\documentclass[fleqn,twocolumn,tighten,apj,twocolappendix]{aastex63} 
	
\usepackage[english]{babel}
\usepackage{blindtext}
\usepackage{CJKutf8}

\usepackage{amsmath,bm,nccmath,textcomp,gensymb}

\usepackage{scalerel}

\usepackage{hyperref,multirow,dsfont,balance,enumitem,tabularx,booktabs}

\usepackage[titletoc]{appendix}

\newcommand\FeI{\ion{Fe}{1}}

\newcommand\NiI{\ion{Ni}{1}}
\newcommand\TiI{\ion{Ti}{1}}
\newcommand\SI{\ion{S}{1}}
\newcommand\SiI{\ion{Si}{1}}
\newcommand\NaI{\ion{Na}{1}}

\newcommand{\aks}[1]{{\color{black}{#1}}} 

\shorttitle{MADGICS 8621\,\r{A} DIB Catalog}
\shortauthors{Saydjari et al.}

\graphicspath{{./}{figures/}}

\begin{document}

\title{Measuring the 8621\,\r{A} Diffuse Interstellar Band in Gaia DR3 RVS Spectra: \\
Obtaining a Clean Catalog by Marginalizing over Stellar Types}

\correspondingauthor{Andrew Saydjari}
\email{andrew.saydjari@cfa.harvard.edu}

\author[0000-0002-6561-9002]{Andrew K. Saydjari}
\affiliation{Department of Physics, Harvard University, 17 Oxford St., Cambridge, MA 02138, USA}
\affiliation{Harvard-Smithsonian Center for Astrophysics, 60 Garden St., Cambridge, MA 02138, USA}

\author[0000-0001-9308-0449]{Ana Sofía M. Uzsoy}
\affiliation{Harvard-Smithsonian Center for Astrophysics, 60 Garden St., Cambridge, MA 02138, USA}

\author[0000-0002-2250-730X]{Catherine Zucker}
\altaffiliation{Hubble Fellow}
\affiliation{Harvard-Smithsonian Center for Astrophysics, 60 Garden St., Cambridge, MA 02138, USA}
\affiliation{Space Telescope Science Institute, 3700 San Martin Drive, Baltimore, MD 21218, USA}

\author[0000-0003-4797-7030]{J. E. G. Peek}
\affiliation{Space Telescope Science Institute, 3700 San Martin Drive, Baltimore, MD 21218, USA}
\affiliation{Department of Physics \& Astronomy, Johns Hopkins University, Baltimore, MD 21218, USA}

\author[0000-0003-2808-275X]{Douglas P. Finkbeiner}
\affiliation{Department of Physics, Harvard University, 17 Oxford St., Cambridge, MA 02138, USA}
\affiliation{Harvard-Smithsonian Center for Astrophysics, 60 Garden St., Cambridge, MA 02138, USA}

\begin{abstract}
Diffuse interstellar bands (DIBs) are broad absorption features associated with interstellar dust and can serve as chemical and kinematic tracers. Conventional measurements of DIBs in stellar spectra are complicated by residuals between observations and best-fit stellar models. To overcome this, we simultaneously model the spectrum as a combination of stellar, dust, and residual components, with full posteriors on the joint distribution of the components. This decomposition is obtained by modeling each component as a draw from a high-dimensional Gaussian distribution in the data-space (the observed spectrum)---a method we call ``Marginalized Analytic Data-space Gaussian Inference for Component Separation'' (MADGICS). We use a data-driven prior for the stellar component, which avoids missing stellar features not well-modeled by synthetic spectra. This technique provides statistically rigorous uncertainties and detection thresholds, which are required to work in the low signal-to-noise regime that is commonplace for dusty lines of sight. We reprocess all public Gaia DR3 RVS spectra and present an improved 8621\,\r{A} DIB catalog, free of detectable stellar line contamination. We constrain the rest-frame wavelength to $8623.14 \pm 0.087$ \r{A} (vacuum), find no significant evidence for DIBs in the Local Bubble from the $1/6^{\rm{th}}$ of RVS spectra that are public, and show unprecedented correlation with kinematic substructure in Galactic CO maps. We validate the catalog, its reported uncertainties, and biases using synthetic injection tests. We believe MADGICS provides a viable path forward for large-scale spectral line measurements in the presence of complex spectral contamination.
\end{abstract}

\keywords{Diffuse interstellar bands (379), Astronomy data reduction (1861), Catalogs (205)}

\section{Introduction} \label{sec:Intro}

Diffuse interstellar bands (DIBs) are broad absorption features associated with interstellar dust and can be used as chemical tracers, though few DIB carriers have been conclusively identified \citep{Campbell_2015_Natur}. Despite this, the moments of a simple Gaussian model of the DIB line shape contain a wealth of information. Comparisons of the $0^{\rm{th}}$ moment (the integral under the curve, often called the equivalent width, $EW_{\rm{DIB}}$), with dust extinction carries information about variations in the \aks{interstellar medium (ISM)} environment which may enhance or deplete DIB carriers \citep{Lan_2015_MNRAS}. The $1^{\rm{st}}$ moment (the detection wavelength, $\lambda_{\rm{DIB}}$), can be converted into a velocity ($v_{\rm{DIB}}^{\rm{LSR}}$) given a known rest-frame wavelength and used as a kinematic tracer of the ISM. Previous work used the 15273 \r{A} \aks{near-infrared (NIR)} DIB detected with APOGEE \citep{Majewski_2017_AJ} to map the ISM velocity field and measure large-scale Galactic rotation \citep{Zasowski_2015_ApJ} and spiral substructure \citep{Tchernyshyov_2018_AJ}. The $2^{\rm{nd}}$ moment (the width of the Gaussian, $\sigma_{\rm{DIB}}$), contains information about the velocity distribution of dust along the line of sight, the instrumental line spread function, and the intrinsic lifetime of the DIB carrier, useful for carrier identification \citep{Campbell_2015_Natur}. While \aks{coarse estimates of} of $\sigma_{\rm{DIB}}$ \aks{have} been used in the past, careful reporting of single-detection $\sigma_{\rm{DIB}}$ and corresponding uncertainties remains a challenge.

Early identification of DIBs focused on taking high SNR, and sometimes high resolution, spectra of a handful of stars along lines of sight with high extinction \citep{vanLoon_2013_A_A}. Features that correlated with dust extinction and were not present in low-extinction reference stars were identified as DIBs. This often limited those studies to hot stars with relatively few stellar features in order to minimize confusion between the DIBs and stellar features. More recently, larger population studies in the Sloan Digital Sky Survey (SDSS, \citealt{York_2000_AJ}) and the Radial Velocity Experiment (RAVE, \citealt{Steinmetz_2006_AJ}) have utilized stacking in order to improve SNR enough for statistically significant detections \citep{Kos_2013_ApJ,Lan_2015_MNRAS}. While stacking can enable detections of fainter DIBs, the information contained in the moments of the Gaussian profile of the DIBs is significantly reduced in that they are averaged over all of the combined spectra.

Now, sufficient data quality and quantities are available to enable large-scale single detection DIB catalogs. In producing a catalog, we produce a low-dimensional representation of the detected DIBs and their properties, generally by listing their 0$^{\rm{th}}$, 1$^{\rm{st}}$, and 2$^{\rm{nd}}$ moments along with the stars toward which they were detected. This has been done in the NIR for the 15273 \r{A} DIB in APOGEE \citep{Zasowski_2015_ApJ}, which searched over $\sim 100$k spectra, and in the optical for the 8621\,\r{A} DIB in Gaia RVS spectra \citep{Schultheis_2022_arXiv}, which searched over $\sim 5.5$M spectra. Yet, statistical methods for modeling and detecting DIBs have not kept pace with the increase in data quantity and quality. 

DIB catalogs are still made by obtaining a best fit stellar spectrum, dividing the observed spectrum by the stellar model, and fitting a Gaussian peak to the residuals centered near the DIB wavelength \citep{Zasowski_2015_ApJ,Schultheis_2022_arXiv}. However, stellar line lists and models often have significant residuals relative to observations, which can cause confusion when fitting or finding DIB features in the residuals. Further, such fits are susceptible to biases resulting from variability in the continuum normalization and are often limited to the high stellar SNR regime, despite the fact that dusty lines of sight are more likely to have low stellar SNR due to dust extinction.

In this work, we introduce a new method for detecting and modeling DIBs that jointly models the stellar and DIB components of the spectra, marginalizing over stellar types and uncertainties in stellar lines. This method is termed ``Marginalized Analytic Data-space Gaussian Inference for Component Separation'' (MADGICS). We reprocess all public Gaia DR3 RVS spectra with MADGICS and demonstrate reduced stellar line contamination in the new MADGICS DIB catalog presented herein. We believe that the speed and stability of the method, even at low stellar SNR, establishes MADGICS as an important statistical tool in the era of large spectroscopic (DIB) surveys. We anticipate the applicability of this method also extends to atomic interstellar absorption \citep{Welty_2001_ApJS} and emission \citep{Kollmeier_2017_arXiv} line surveys. 

We begin by comparing the 8621\,\r{A} MADGICS DIB catalog to the Gaia DIB catalog and presenting astrophysical applications and validations of the catalog (Section \ref{sec:Astro}). Then, we describe the method, catalog construction, and further validation tests (Sections \ref{sec:Model}--\ref{sec:Build}).


\section{Data} \label{sec:Data}

The primary data used in this work are the Gaia DR3 RVS spectra, which were obtained from the Gaia archive \citep{GaiaCollaboration_2016_A_A,Vallenari_2022_arXiv}. These are optical spectra spanning $8460-8700$ \r{A}, which are shifted into the stellar rest frame, averaged over all visits, and interpolated onto a common linear (0.1 \r{A} spacing) wavelength grid (2401 bins). The wavelength calibration of the spectra are reported in vacuum.\footnote{Throughout this work, we work exclusively in vacuum wavelength, except in referring to the catalog and DIB by the common 8621\,\r{A} nomenclature reflective of the historical identification of DIBs by wavelength in air.} Both the flux and flux uncertainty per wavelength bin are provided by Gaia. See Section 5.2 of \citealt{Vallenari_2022_arXiv} for more details.\vspace{-3 mm}

\begin{deluxetable}{cccc}[h]
\tablenum{1}
\tablecaption{8621\,\r{A} DIB Catalogs
\label{tab:dibcat}}
\tablecolumns{4}
\tablehead{
Cuts & MADGICS & Gaia HQ & Gaia Full
}
\startdata
\multicolumn{4}{c}{\emph{SNR Cuts}}\\
\hline
Stellar SNR & $>15$ &  $>70$ &  $>70$ \\ 
DIB SNR & $ > 3.8$ &  $> 2.86$ &  \\ 
\hline
\multicolumn{4}{c}{\emph{Overall Spectrum Goodness of Fit Cuts}}\\
\hline
$\chi^2_{\rm{tot}}/\rm{dof}$ & $0.71-1.41$ & &  \\ 
GSP-Spec Flags & & $< 2$ &  \\ 
$\alpha_{\rm{DIB}}$ & & $\alpha_{\rm{DIB}} > R_i$ & \\
$\lambda_{\rm{DIB}}$ & & $8620-8626$ \r{A} &  \\ 
$\sigma_{\rm{DIB}}$ & & $0.6-3.2$ \r{A} &  \\
\hline
\multicolumn{4}{c}{\emph{Number of Spectra with Detections}}\\
\hline
Public (999,645) & 7,789 & 9,763 & 50,787 \\ 
Total (5,591,594) & & 137,536 & 476,532 \\ 
\enddata
\end{deluxetable}\vspace{-12mm}

The primary Gaia analysis of these spectra falls under the General Stellar Parameteriser-spectroscopy (GSP-Spec) module, which is a purely spectroscopic analysis to obtain chemo-physical stellar parameters from combined RVS spectra \aks{\citep{Recio-Blanco_2022_arXiv}}. Within GSP-Spec, DIB detections and parameter estimates were made predominantly by dividing each RVS spectrum by the best matching synthetic spectrum (using the GSP-Spec atmospheric parameters) and performing a least-squares fit of a Gaussian function \citep{Schultheis_2022_arXiv}. 

\begin{figure*}[t]
\centering
\includegraphics[width=\linewidth]{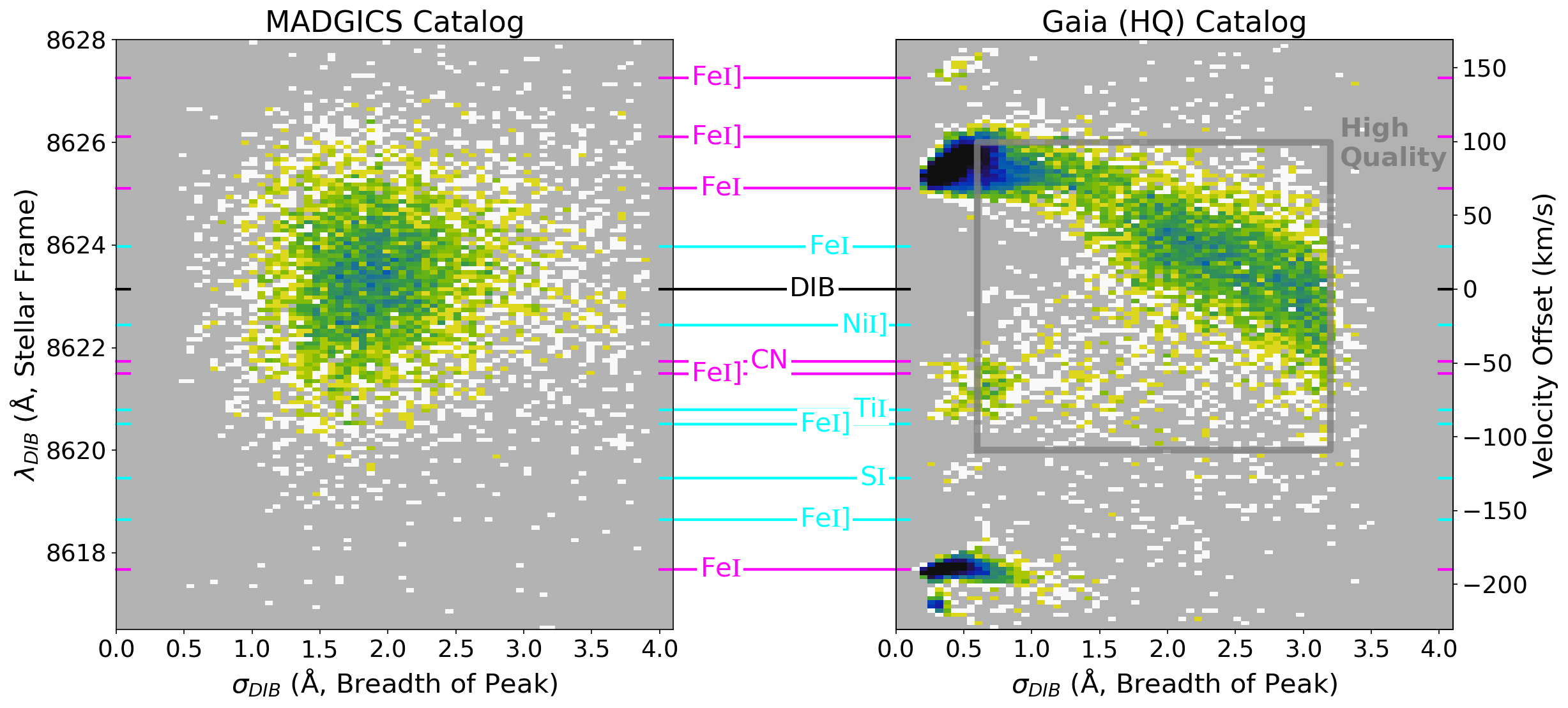}
\caption{Density of detections in the MADGICS (this work, \textit{left}) and Gaia \citep[][\textit{right}]{Schultheis_2022_arXiv} 8621\,\r{A} DIB catalog as a function of the DIB width ($\sigma_{\rm{DIB}}$) and its central wavelength ($\lambda_{\rm{DIB}}$). The color scale indicates density, where darker colors indicate more detections, and is a common logarithmic stretch for both catalogs. The ($\lambda_{\rm{DIB}}$, $\sigma_{\rm{DIB}}$) cuts imposed in defining the Gaia HQ sample are shown as a gray box, inside of which the HQ sample is shown. Outside of that box, the Gaia catalog is shown with only the stellar SNR, DIB SNR, and GSP-Spec flag cuts imposed. Guidelines for the central wavelength of the DIB and stellar lines are labeled and color coded between the plots. Stellar lines in magenta indicate lines not included or not well-calibrated in the GSP-Spec line list, and the Gaia DIB detections preferentially cluster at those wavelengths, especially at low $\sigma_{\rm{DIB}}$. Because the MADGICS catalog inference marginalizes over a data-driven stellar model, it is free from detectable contamination from these stellar lines and provides a more robust characterization of the DIB properties.}
\label{fig:contamfig}
\end{figure*}

Of the 5,591,594 RVS spectra processed by GSP-Spec, only 999,645 are publicly available. Based on the full 5,591,594 spectra, the Gaia DIB catalog has entries for 476,532 spectra, of which 50,787 have public spectra (Table \ref{tab:dibcat}). The Gaia DIB team strongly suggests restricting this catalog to a ``high quality'' (HQ) sample, which includes 137,536 spectra, of which 9,763 are public \citep{Schultheis_2022_arXiv}. While the Gaia DIB catalog entries are available for all stars---even those without public spectra---we will restrict our comparisons with the Gaia DIB catalog to stars with public spectra, unless otherwise specified. The Gaia DIB catalog, and other Gaia-derived properties of the stars or spectral processing were obtained from the DR3 ``source'' and ``astrophysical parameters'' tables.

The cuts defining the Gaia HQ sample are summarized in Table \ref{tab:dibcat}. We divide these cuts into two classes: cuts on (1) SNR and (2) the overall model goodness of fit for the spectrum. The entire Gaia 8621\,\r{A} DIB catalog is limited to stellar SNR $> 70$. A further SNR cut on the DIB detection is imposed by requiring $EW_{\rm{DIB}}/\sigma(EW_{\rm{DIB}}) > 2.86$. The efficacy of cuts imposed this way of course depends strongly on the quality of the $\sigma(EW_{\rm{DIB}})$ estimate. 

Additional cuts \citep{Schultheis_2022_arXiv} are imposed to remove low-quality detections, often as a result of poor stellar modeling. The GSP-Spec flags 1-13, which indicate the goodness of fit for the stellar modeling, are required to be $< 2$ (reasonably good). Cuts on the DIB peak fit parameters limit $\lambda_{\rm{DIB}}$ and $\sigma_{\rm{DIB}}$ to reasonable ranges, $8620-8626$ \r{A} and $0.6-3.2$ \r{A}, respectively. The amplitude of the DIB peak, $\alpha_{\rm{DIB}}$, is required to be larger than a residual threshold $R_i$, which has a constant threshold of at least $0.15$. The residual threshold $R_i$ depends on $\sigma_{\rm{DIB}}$ and is max($R_A,R_B,0.15$) for $0.6 < \sigma_{\rm{DIB}} < 1.2$ and max($R_B,0.15$) for $1.2 < \sigma_{\rm{DIB}} < 3.2$. Here $R_A$ denotes the standard deviation of the ``global'' data-model residuals 8605-8640 \r{A} and $R_B$ denotes the standard deviation of the ``local'' residuals ``within the DIB profile,'' neither of which are publicly available \citep{Schultheis_2022_arXiv}. 

The selection function of public RVS spectra remains forthcoming \citep{Seakbrokeunpub}, but is nonuniform across the sky and generally concentrated within 1-2 kpc of the Sun (see Appendix \ref{sec:selectFun}). There is also significant heterogeneity in the normalization of the RVS spectra (see Appendix \ref{sec:DIBSNR}) as a result of a discrete choice in the Gaia pipeline to either pseudo-continuum normalize or rescale the spectra by a constant, as in the case of cool or low SNR stars. \citep{Vallenari_2022_arXiv}. \vspace{-0.5em}


\section{Astrophysical Catalog Validation} \label{sec:Astro}

\subsection{Stellar Frame} \label{sec:StellarFrame}

We first compare the MADGICS and Gaia HQ 8621\,\r{A} DIB catalog in terms of the $1^{\rm{st}}$ and $2^{\rm{nd}}$ moments of the detected peak---the center wavelength in the stellar frame $\lambda_{\rm{DIB}}$ and its width $\sigma_{\rm{DIB}}$ (Figure \ref{fig:contamfig}). The advantage of plotting the catalog in the stellar frame is that it readily identifies the impact of stellar features on the DIB identification. In both panels, we provide guidelines corresponding to the:
\vspace{-0.5em}
\begin{itemize}
    \setlength\itemsep{-0.5em}
    \item putative DIB rest-frame wavelength $\lambda_{\rm{DIB_{rest}}}$ (black),
    \item calibrated stellar lines in the GSP-Spec line list (turquoise, \citealt{Contursi_2021_A_A,vizier}),
    \item and stellar lines not included or not well-calibrated in GSP-Spec, but coincident with features of stellar origin (see also Section \ref{sec:Prior}) (magenta).
\end{itemize}\vspace{-2 mm}
For the lines not well-modeled by GSP-Spec, we do not claim or validate assignment of the observed stellar features as resultant from these lines. We simply plot the wavelength and identification of the nearest transition with a large contribution to the solar spectrum from the Kurucz archive (\href{http://kurucz.harvard.edu/}{http://kurucz.harvard.edu/}) (see Section \ref{sec:stellarprior} and Appendix \ref{sec:linelist}). We focus on demonstrating the impact of stellar lines on DIB catalogs rather than an identification of the stellar lines themselves.

In the right panel, the Gaia HQ catalog is shown within the central gray box, which denotes the $(\lambda_{\rm{DIB}}, \sigma_{\rm{DIB}})$ cuts imposed in defining the Gaia HQ sample (see Table \ref{tab:dibcat}). Outside of that box, we show DIB detections from the broader Gaia DIB catalog with only the stellar SNR, DIB SNR, and GSP-Spec flag cuts imposed. Even inside the HQ catalog limits (gray box), the influence of stellar features is apparent. The largest density of detections peaks at the edge of the $\sigma_{\rm{DIB}}$ cut (0.6 \r{A}) and is centered on the 8625.11 \r{A} \FeI \ line, that is not in the GSP-Spec line list (see Appendix \ref{sec:linelist}). This peak in density extends to larger $\sigma_{\rm{DIB}}$ until $\sim 1.5$ \r{A} when it broadens and shifts slowly to shorter wavelengths. While not clearly associated with a specific stellar line above $\sigma_{\rm{DIB}} = 1.5$ \r{A}, the correlation between $\sigma_{\rm{DIB}}$ and $\lambda_{\rm{DIB}}$ suggests a non-negligible impact of the stellar lines on the properties of the DIB detection. There is also a second peak at low $\sigma_{\rm{DIB}}$ which appears to straddle the 8620.51 \r{A} \FeI],\footnote{The bracket is spectroscopic notation indicating that the transition is an intercombination line, meaning it is semi-forbidden.} 8620.79 \r{A} \TiI, 8621.49 \r{A} \FeI] and 8621.73 \r{A} $^{\bullet}$CN lines. 

These correlations continue and become more prominent outside the parameter cuts (gray box) included in the definition of the HQ sample. An additional peak in the detection density appears near the wavelength of the 8617.67 \r{A} \FeI \ line, that is also not in the GSP-Spec line list (see Appendix \ref{sec:linelist}). The broad 8620-8622 \aks{\r{A}} peak in density splits at lower $\sigma_{\rm{DIB}}$ into two modes centered on either the 8620.51 \r{A} \FeI] \ and 8620.79 \r{A} \TiI \ lines or the 8621.49 \r{A} \FeI] and 8621.73 \r{A} $^{\bullet}$CN lines. Peaks in density near other GSP-Spec stellar lines appear to a lesser extent. 

To contextualize the impact of these wavelength pileups on kinematic DIB studies, the right y-axis shows the radial velocity corresponding to the wavelength shift relative to the DIB center wavelength. This is not indicative of the true DIB radial velocity, which must account for the stellar radial velocity (see Section \ref{sec:Kin}).

\begin{figure}[b]
\centering
\includegraphics[width=\linewidth]{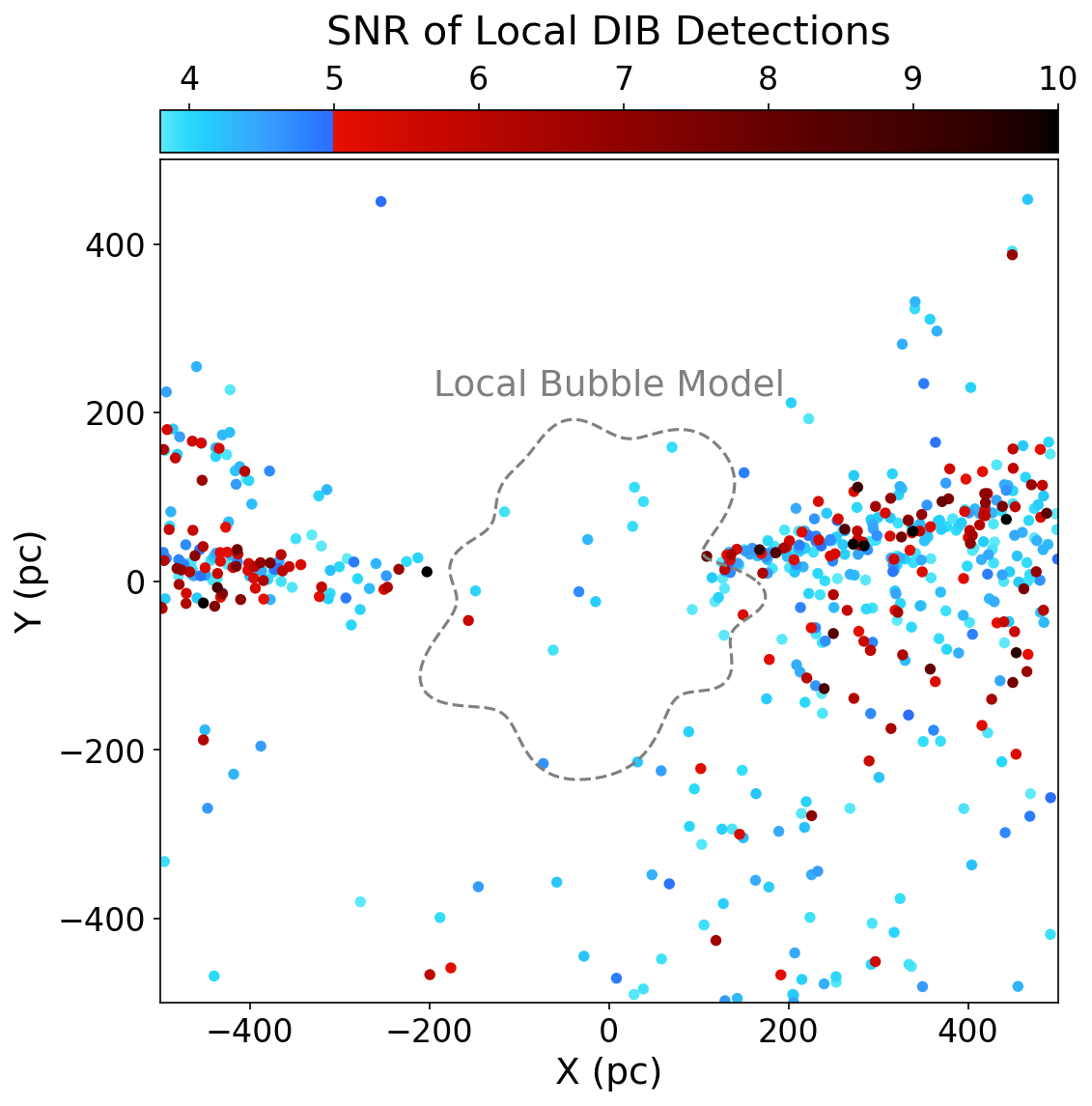}
\caption{MADGICS DIB detections within the Local Bubble, where color indicates the SNR of the detections. A 2D slice at the midplane (Z=0 pc) of the boundary for a model of the Local Bubble \citep{Pelgrims_2020_A,DVN/RHPVNC_2022} is shown (dashed gray). In the MADGICS catalog, no confident detections (DIB SNR $> 5$) are present within the Local Bubble model. This is in contrast to previous work, though we find evidence that previously reported detections were associated with stellar features (Figure \ref{fig:contamfig}). \aks{An interactive version of this figure is available \href{https://faun.rc.fas.harvard.edu/czucker/Paper\_Figures/DIB\_Local\_Bubble\_SNR\_3D.html}{here}. The interactive version provides a manipulable 3D view of the DIB detections, Local Bubble model, location of the Sun, and estimates of the dust distribution from \cite{Leike_2020_yCat}, with the ability to turn each layer on/off individually. Hover tools over each DIB detection also provide the DIB SNR, $EW_{\rm{DIB}}$, $v_{\rm{DIB}}^{\rm{LSR}}$, and $\sigma_{\rm{DIB}}$.}}
\label{fig:localbubble}
\end{figure}

This comparison with the less restrictive sample \aks{suggests} that the Gaia goodness of fit cuts on $(\alpha_{\rm{DIB}}, \lambda_{\rm{DIB}}, \sigma_{\rm{DIB}})$ are not fully effective at removing the influence of stellar lines on the DIB catalog, even within the HQ sample. This list of stellar features are either poorly-modeled by or missing from GSP-Spec that impact the Gaia 8623\r{A} DIB catalog is not meant to be exhaustive, but merely illustrative that detections of unmodeled stellar features dominate the catalog, with a small contribution from even the carefully calibrated stellar features known in GSP-Spec as well (see Appendix \ref{sec:linelist}). A comparison of both the MADGICS and Gaia catalogs \aks{removing their respective} goodness of fit cuts is provided in Appendix \ref{sec:noGOF}. In general, we aim to (1) impose goodness of fit cuts that eliminate any influence of stellar lines and (2) use methods that are maximally robust to stellar mismodeling so that real DIBs can be detected and modeled well in the largest number of spectra possible. \aks{These quality cuts for the MADGICS DIB catalog are detailed in Table \ref{tab:dibcat}.}

\begin{figure*}[t]
\centering
\includegraphics[width=\linewidth]{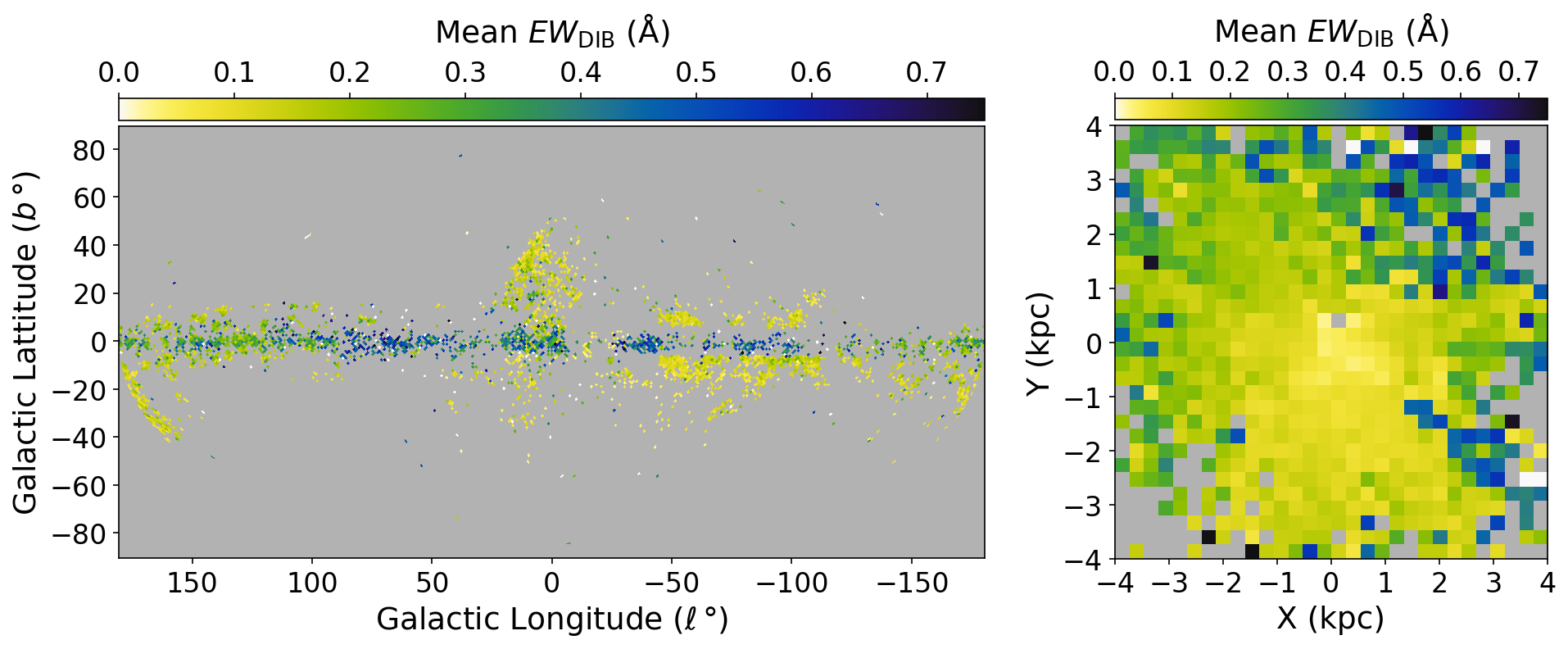}
\caption{Projections of the 3D DIB EW map in (left) angular ($\ell , b$) and (right) Cartesian ($X, Y$) Galactic coordinates. The inverse-variance weighted mean of $EW_{\rm{DIB}}$ is shown in color, averaging over dimensions not shown.}
\label{fig:sky_ew}
\end{figure*}

In the left panel, the MADGICS catalog morphology is dominated by a Gaussian centered at $(\lambda_{\rm{DIB}}, \sigma_{\rm{DIB}}) = (8623.37$ \r{A}, $1.9$ \r{A}$)$. It has a slightly heavy tail toward larger $\sigma_{\rm{DIB}}$, which can be understood in terms of a low-DIB-SNR detection bias (see Section \ref{sec:LowSNR}). The $\lambda_{\rm{DIB}}$ dispersion is 1.3 \r{A}. Assuming all of the $\lambda_{\rm{DIB}}$ dispersion is the result of the relative star-dust velocity dispersion, this $\lambda_{\rm{DIB}}$ dispersion corresponds to a velocity dispersion of $46$ km/s. This velocity dispersion is consistent with a value of $49$ km/s, which is obtained by adding in quadrature the dispersion of radial velocities for the stars in the sample measured by Gaia (35 km/s) and the typical velocity dispersion of CO gas ($\sim 35$ km/s, as a proxy for the dust velocity) \citep{Dame_2001_ApJ}. We explicitly note there is no variation in the density of detections or the $\sigma_{\rm{DIB}}$ as the $\lambda_{\rm{DIB}}$ approaches any of the stellar features indicated.

Thus, the MADGICS catalog delivers a distribution of DIB detections, free of detectable contamination by stellar lines, with a catalog defined only by simple SNR-based detection thresholds and a single $\chi^2$ cut representing the overall goodness of fit. This constitutes a clear improvement over \aks{conventional methods of least-squares fitting DIB line shape functions to the data divided by a best-fit synthetic stellar model.}

\subsection{Local Bubble} \label{sec:LocalBubble}

While DIBs are known to be correlated with dust extinction, their carriers are suspected to be similar to the only clearly identified DIB carrier, the buckminsterfullerene cation (C$_{60}^+$) \citep{Edwards_1993_A_A,Campbell_2015_Natur,Walker_2015_ApJL}. Thus, DIBs should be detected when large molecules are present, not dust grains themselves, and the correlation is likely due to coupled formation mechanisms. A recent body of literature has claimed detections of DIBs within the Local Bubble, despite the low dust extinction therein \citep{Bailey_2016_A_A,Farhang_2019_NatAs,Schultheis_2022_arXiv}. The Gaia HQ sample supports these claims by reporting confident detections, albeit of low EW, within the Local Bubble.\footnote{A large number of such detections are included in the public subset of Gaia RVS spectra.}

Within the MADGICS DIB catalog, there are no confident DIB detections (DIB SNR $> 5$) within the Local Bubble (Figure \ref{fig:localbubble}). All confident DIB detections fall outside the bubble and cluster along lines of sight passing through known local molecular clouds such as Taurus, Perseus, and Ophiuchus \citep{Zucker_2022_Natur}. In projection, three significant detections appear to fall inside the bubble, but do not when examined in 3D. The 3D distribution can be visualized with the \href{https://faun.rc.fas.harvard.edu/czucker/Paper\_Figures/DIB\_Local\_Bubble\_SNR\_3D.html}{interactive version} of Figure \ref{fig:localbubble}. This is an important demonstration of the necessity of marginalizing over possible stellar models when reporting the confidence of DIB detections. Using a single fixed stellar model per star can easily lead to stellar residuals that can confuse conventional DIB fitting approaches because there is confidently a feature in the stellar residuals, but that feature is of stellar origin. The MADGICS DIB SNR is instead based upon the $\Delta\chi^2$ resulting from adding a DIB component to the component separation, using an empirical model to marginalize over stellar features (see Section \ref{sec:Model}).

Because the MADGICS stellar model is empirical (see Section \ref{sec:Prior}), one might worry that, if they exist, low EW DIBs uncorrelated with dust are baked into the stellar model. While possible, we show in Section \ref{sec:Prior} no features resembling the DIB in the stellar component prior. Further, upon loosening the cut on training stars used in constructing the stellar prior (allowing dustier stars into the training sample), we find only slightly reduced sensitivity to DIB detection (See Appendix \ref{sec:DustyPrior}). Thus, using a more careful definition of DIB SNR, we find no evidence for DIB detections within the Local Bubble in public Gaia DR3 RVS spectra \citep{Schultheis_2022_arXiv}, which calls for careful reconsideration of related previous claims. We look forward to the future release of more and higher stellar SNR spectra to help resolve this question. \vspace{-3 mm}

\subsection{Moment-0} \label{sec:Mom0}

\begin{figure}[t]
\centering
\includegraphics[width=\linewidth]{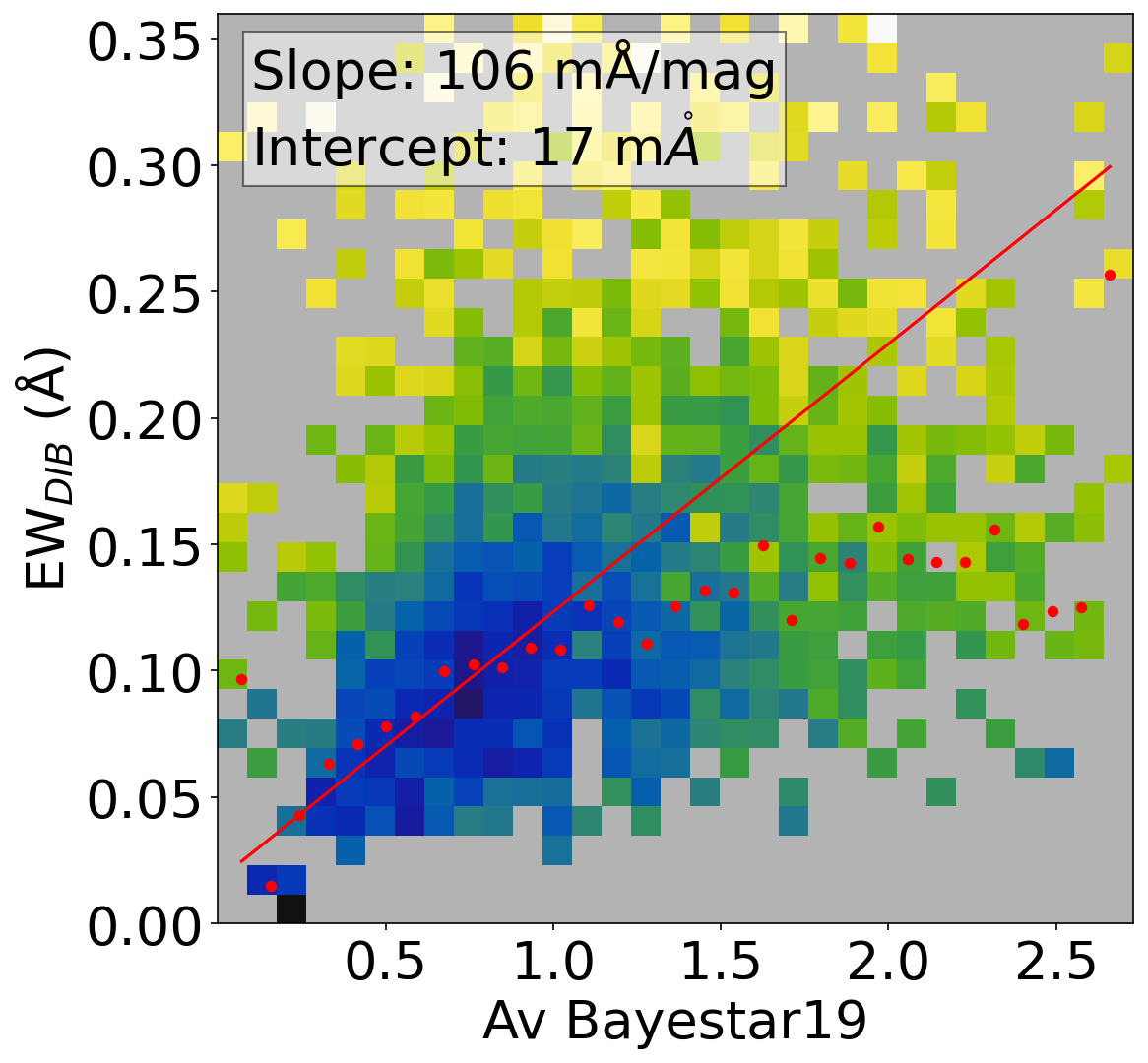}
\caption{Weighted least-squares linear fit of $EW_{\rm{DIB}}$ versus extinction to the background star as measured by Bayestar19 \citep{Green:2019:ApJ:} in the range $0 < A_V < 1$. The color scale indicates $EW_{\rm{DIB}}$ inverse variance weighted number density, where darker colors indicate more (confident) detections and a log normalization has been applied. Scatter points represent inverse-variance weighted mean $EW_{\rm{DIB}}$ for each Av bin.}
\label{fig:ew_linear}
\end{figure}

The moment-0 ($EW_{\rm{DIB}}$) map of DIB detections (Figure \ref{fig:sky_ew}) closely resembles a noisy version of 3D dust maps obtained using other techniques \citep{Green:2019:ApJ:}. Figure \ref{fig:sky_ew} also clearly features detections predominantly along lines of sight with the highest extinction. This makes sense given the known positive correlation between DIBs and dust and the low to moderate SNR of most of the stellar spectra; confident DIB detections (high DIB SNR) require large DIB signals and, by correlation, large dust column densities.

\begin{figure*}[t]
\centering
\includegraphics[width=\linewidth]{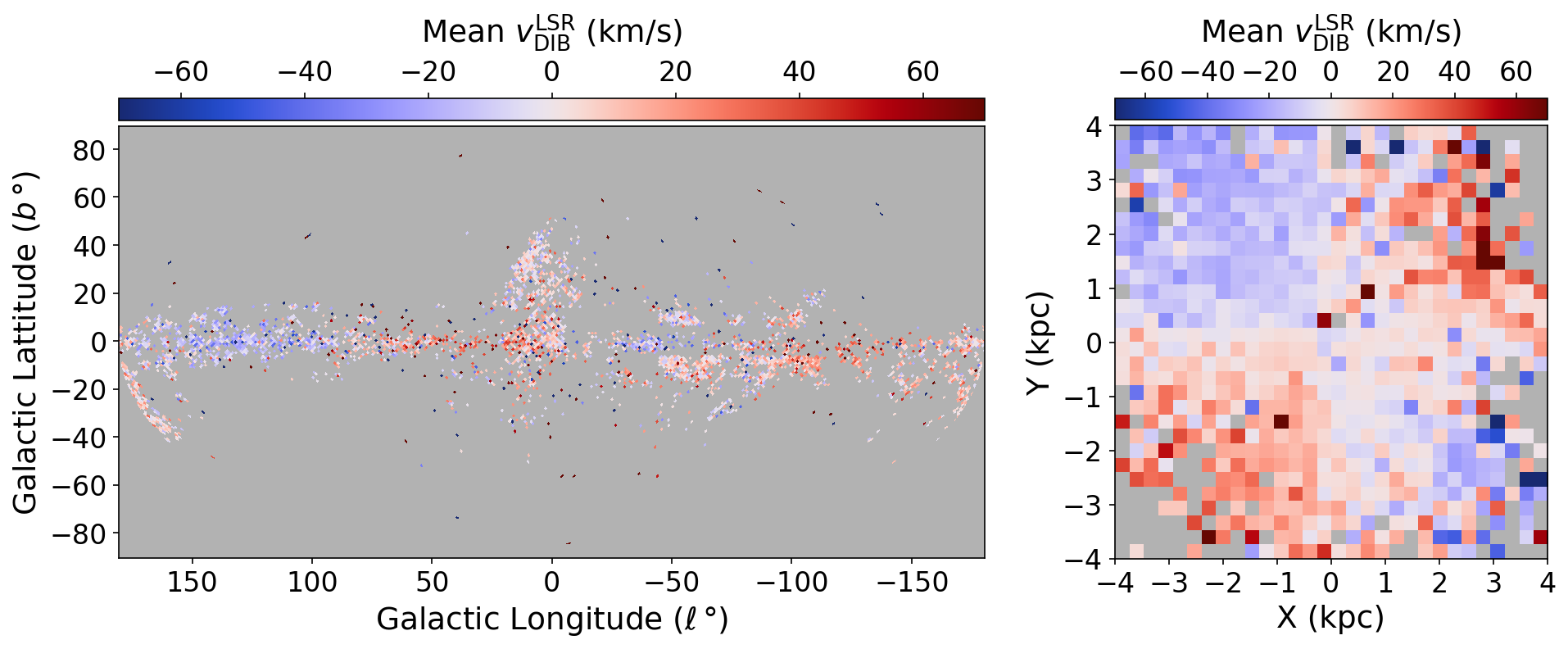}
\caption{Projections of the 4D DIB radial velocity map in (left) angular ($\ell , b$) and (right) Cartesian ($X, Y$) Galactic coordinates. The inverse-variance weighted mean of $v_{\rm{DIB}}^{\rm{LSR}}$ is shown in color, averaging over dimensions not shown.}
\label{fig:sky_vel}
\end{figure*}

\begin{figure*}[t]
\centering
\includegraphics[width=\linewidth]{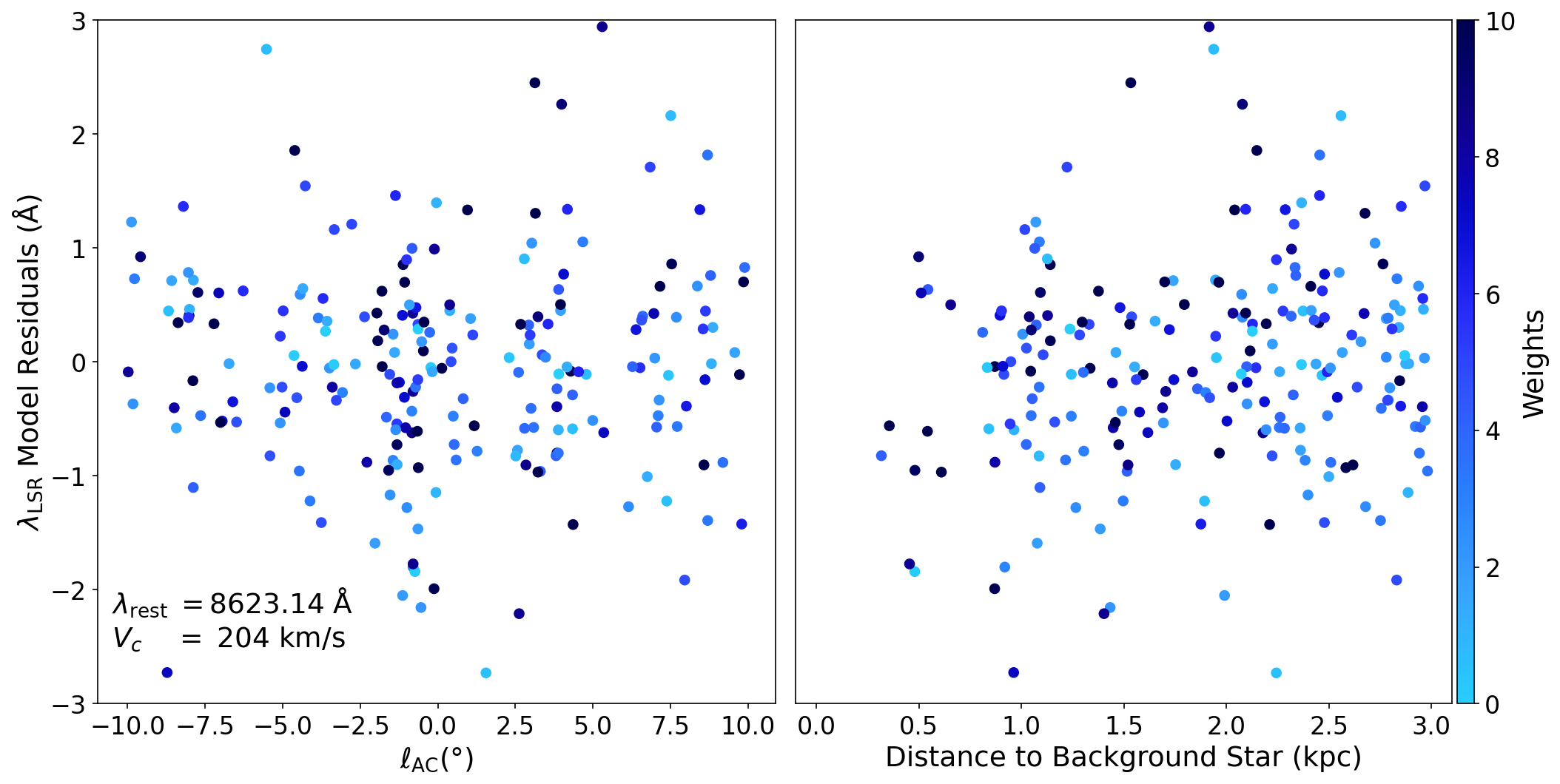}
\caption{Scatter plot showing residuals after fitting $\lambda_{\rm{LSR}}$ versus $\ell'$ and $d$ where the color shows the weighting of each point in the fit. Only detections such that $|\ell'| < 10 \degree$, $|b| < 10 \degree$, and $d < 3$ kpc are shown. Our derived rest-frame wavelength of $\lambda_{\rm{rest}} = 8623.14 \pm 0.087$ \r{A} is bluer than \cite{Schultheis_2022_arXiv}, but our updated value is validated in Figure \ref{fig:dust_co_corr} by comparison with CO gas velocities.}
\label{fig:lam_fit}
\end{figure*}

Studies of DIBs often focus on the linearity of the correlation of the $EW_{\rm{DIB}}$ with dust column density, extinction $A_{V}$, or reddening $E(B-V)$ \citep{Kos_2013_ApJ,Schultheis_2022_arXiv}. However, several DIBs are known to ``flatten-out'' or even ``turn-over'', having a negative correlation of $EW_{\rm{DIB}}$ with dust extinction past a critical point \citep{Lan_2015_MNRAS}. This could be, for example, because the DIB carrier freezes out of the gas phase onto grains, chemical reactions deplete the DIB carrier, or there are several dust clouds along the line of sight, some of which do not host DIBs.

Previous work using purely statistical correlations, without imposing detection cuts, have found linear correlations with near zero intercepts \citep{Zasowski_2015_ApJ}. However, the creation of a catalog necessarily imposes a detection bias that more significantly impacts DIBs at low reddening and can thus lead to a positive intercept of $EW_{\rm{DIB}}$ with $A_{V}$.

In Figure \ref{fig:ew_linear}, we show a weighted linear least-squares fit of $EW_{\rm{DIB}}$ versus the extinction to the background star, as measured by the Bayestar19 3D dust map \citep{Green:2019:ApJ:}. The weights are simply the inverse-variance of the $EW_{\rm{DIB}}$ and the fit is performed in the range $0 < A_V < 1$. We find a slope of $106\pm33$ m\r{A}/mag and an intercept of $17\pm26$ m\r{A}. 

Because Bayestar19 has significant scatter around the line of sight fits and we do not know what fraction of the reddening to the star is associated with the DIB, there is a large, but hard to quantify, uncertainty in the extinction axis. Thus, we derive our error bars on the slope and intercept from independent fits to the DIB detections in each quadrant in Galactic longitude. Further, in all fits and figures, we restrict to lines of sight where the Bayestar19 extinction prediction is unchanged within $\pm1 \sigma$ of the distance to the background star according to the distance uncertainties in \cite{Bailer-Jones_2021_AJ}. While this selection could introduce a bias with respect to Galactic latitude and DIB environment, we leave investigation of these more subtle effects to future work.

For reference, the 15273 \r{A} DIB which has previously been used for successful ISM mapping has a reported slope of $102\pm1$ m\r{A}/mag versus $A_V$ \citep{Zasowski_2015_ApJ}. To directly compare the slope to \cite{Schultheis_2022_arXiv}, which measured the same 8621\,\r{A} DIB, we convert $A_V$ Bayestar19 to $E(B-V)$ to find $290\pm91$ m\r{A}/mag versus $E(B-V)$. While \cite{Schultheis_2022_arXiv} report in mag/m\r{A}, a simple inversion of the slope yields 455 m\r{A}/mag versus $E(B-V)$, which is $1.8\sigma$ larger than the value found here. Careful control of $EW_{\rm{DIB}}$ biases and uncertainties and using those uncertainties in the fit is important to avoid slight overestimation of the slope.
\vspace{-4 mm}
\subsection{Moment-1} \label{sec:Mom1}
\vspace{-1 mm}
\subsubsection{Kinematics} \label{sec:Kin}
\vspace{-1 mm}
To use the DIB to study kinematics (Figure \ref{fig:sky_vel}), we must first convert the $\lambda_{\rm{DIB}}$ measured in the stellar frame to $v_{\rm{DIB}}^{\rm{LSR}}$. We remove the wavelength shift caused by reporting spectra in the stellar frame using the barycentric radial velocity $v_{\rm{star}}$ of the star reported by Gaia.\footnote{We use a positive sign convention for radial velocity moving away from the observer.} At the same time, we remove the wavelength shift caused by the motion of the sun relative to the LSR, given the unit direction vector to the source $\hat{d}$. We use the LSR convention where $\vec{V}_{\rm{Sun}}$ = (U,V,W) = (10.6, 10.7, 7.6) km/s \citep{Reid_2019_ApJ}.\footnote{We use a right-handed convention with U pointing toward the Galactic center and V toward $\ell = 90\degree$.} Then, using the rest-frame wavelength $\lambda_{\rm{rest}}$ and the speed of light $c$, we convert the ``LSR'' wavelength $\lambda_{\rm{LSR}}$ to a velocity.

\begin{ceqn}
\begin{align} \label{eq:wave_lsr}
    \lambda_{\rm{LSR}} &= \lambda_{\rm{DIB}} (1 + v_{\rm{star}}/c + \vec{V}_{\rm{Sun}}/c \cdot \hat{d}) \\
    v_{\rm{DIB}}^{\rm{LSR}} &= \left(1 - \frac{\lambda_{\rm{rest}}}{\lambda_{\rm{LSR}}}\right)c
\end{align}
\end{ceqn}
The reported $v_{\rm{DIB}}^{\rm{LSR}}$ uncertainty is obtained by propagating the $\lambda_{\rm{DIB}}$ uncertainty only and does not include uncertainty contributions from $v_{\rm{star}}$ or overall zeropoint uncertainty related to $\vec{V}_{\rm{Sun}}$.

It is essential to measure the rest-frame (intrinsic) wavelength of the DIB to enable the use of the DIB in studies of Galactic kinematics. Additionally, precise measurement of the rest-frame wavelength aids in identifying its carrier in comparison to theoretical chemical models. Here we use the common method of measuring the wavelength intercept of the DIB detections toward the Galactic anti-center \citep{Zasowski_2015_ApJ,Munari_2008_A_A,Zhao_2021_A_A,Schultheis_2022_arXiv}. Assuming a purely tangential, azimuthally symmetric Galactic rotation, DIB carriers rotating with the Galaxy would exhibit no shift in the $\lambda_{\rm{LSR}}$ relative to the rest-frame wavelength toward the Galactic anti-center.

The expected change in $\lambda_{\rm{LSR}}$ of the DIB as a result of Galactic rotation can be found using Equation \ref{eq:full_gal} when given the distance of the sun to the Galactic center $R_0$, the circular (tangential) velocity of Galactic rotation $V_c$ at the location of the DIB source $\vec{R}$, Galactic longitude of the DIB source referenced to the Galactic anti-center $\ell' = \ell-180$, and distance from the sun to the DIB source projected on to the Galactic plane $d$. This can be linearized in the small angle approximation to yield Equation \ref{eq:small_ang_gal}.

\vspace{-2 mm}
\begin{eqnarray}
    \nonumber \label{eq:full_gal}
    &\frac{\Delta\lambda_{\rm{LSR}}}{\lambda_{\rm{rest}}} = \frac{V_c(\vec{R})\sin{\ell'}}{c} \times \\ 
    &\left[ 1 - \frac{1}{\sqrt{ \sin^2{\ell'} + \frac{d^2}{R_0^2} + 2\frac{d}{R_0}\cos{\ell'} + \cos^2{\ell'}}} \right] \\
    \label{eq:small_ang_gal}
    &\frac{\Delta\lambda_{\rm{LSR}}}{\lambda_{\rm{rest}}} = \frac{V_c(\vec{R})}{c}\frac{d}{d+R_0}\ell' \quad \text{where} \quad \ell' \ll 90\degree
\end{eqnarray}

Because $V_c$ is believed to be locally flat \citep{Bovy_2012_ApJ} we will take $V_c(\vec{R})$ to be a constant. We fix $R_0 = 8.5$ kpc consistent with \cite{Reid_2019_ApJ}. While we only know the distance to the background star, we will use this to approximate $d$, the distance to the DIB source. To help this approximation hold, we limit the fit to background stars within 3 kpc.

In Figure \ref{fig:lam_fit}, we present a least-squares fit of $\lambda_{\rm{LSR}}$ as a function of $\ell'$ and $d$ using Equation \ref{eq:small_ang_gal} and a constant term to measure $\lambda_{\rm{rest}}$. We use inverse-variance weights that sum in quadrature the MADGICS $\sigma(\lambda_{\rm{DIB}})$ and the background star radial velocity uncertainty from Gaia. We restrict to $|\ell'| < 10 \degree$ so that we can safely apply the small angle approximation and $|b| < 10 \degree$ to help approximate the Galactic rotation as uniform in the z-direction. The intercept gives $\lambda_{\rm{rest}} = 8623.14 \pm 0.087$ \r{A} (vacuum). 

To obtain the uncertainty estimate on the least-squares model parameters, we divide the data into $0 < d < 2$ kpc and $2 < d < 3$ kpc. We use the difference in the fit parameters relative to the formal uncertainties to determine a scalar by which to inflate the formal uncertainty so that the difference is only 1 $\sigma$. We perform this procedure to help account for uncertainty in $R_0$, resulting from the approximation of $d$, systematics associated with cuts in $(\ell',b)$ that are imposed, and the rather strong assumption of a purely tangential, azimuthally symmetric Galactic rotation.\footnote{The absolute uncertainty associated with $\vec{V}_{\rm{Sun}}$ always remains.} For example, measurements from \cite{Tchernyshyov_2017_AJ} show a net radial inflow past 1 kpc on the order of $-3$ km/s. We note that the $0.087$ \r{A} uncertainty reported here is comparable to that inflow velocity.

The value found in this work is bluer than that reported in \citealt{Schultheis_2022_arXiv} by 4.7 $\sigma$, using the uncertainties reported therein ($8623.23 \pm 0.019$ \r{A}, vacuum), though only 1.03 $\sigma$ using the uncertainties reported in this work. We further validate the $\lambda_{\rm{rest}}$ and its uncertainty estimate reported here by comparison with $^{12}$CO velocities in Section \ref{sec:12CO}.

An alternative method for determining the rest-frame wavelength is referencing to known atomic ISM transitions, such as \NaI \ \citep{Krelowski_1988_PASP,Galazutdinov_2000_PASP}. While the wavelength of the \NaI \ transition is known with high precision and its use alleviates assumptions about Galactic rotation, surveys measuring both DIBs and transitions for atomic ISM species have historically had far fewer samples. Further, referencing to \NaI \ requires assuming the \NaI \ and DIB carriers are co-moving, which can have large scatter \citep{Farhang_2015_ApJ}.

\begin{figure*}[t]
\centering
\includegraphics[width=\linewidth]{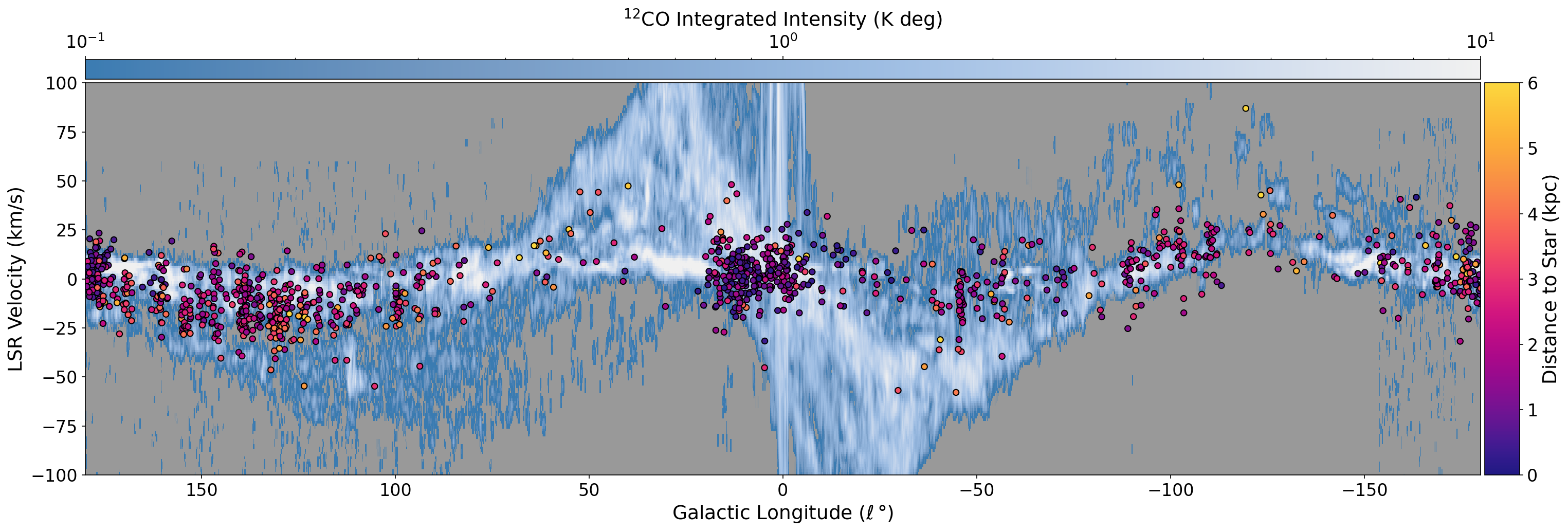}
\caption{Scatter plot of $v_{\rm{DIB}}^{\rm{LSR}}$ for DIB detections with DIB SNR $>5.5$ overlaid on longitude-velocity diagram of CO emission from \citealt{Dame_2001_ApJ}, integrated over $b$ (colorbar top). Scatter points are colored by the distance to the background star (colorbar right).}
\label{fig:VLSR_CO}
\end{figure*}

\begin{figure}[t]
\centering
\includegraphics[width=\linewidth]{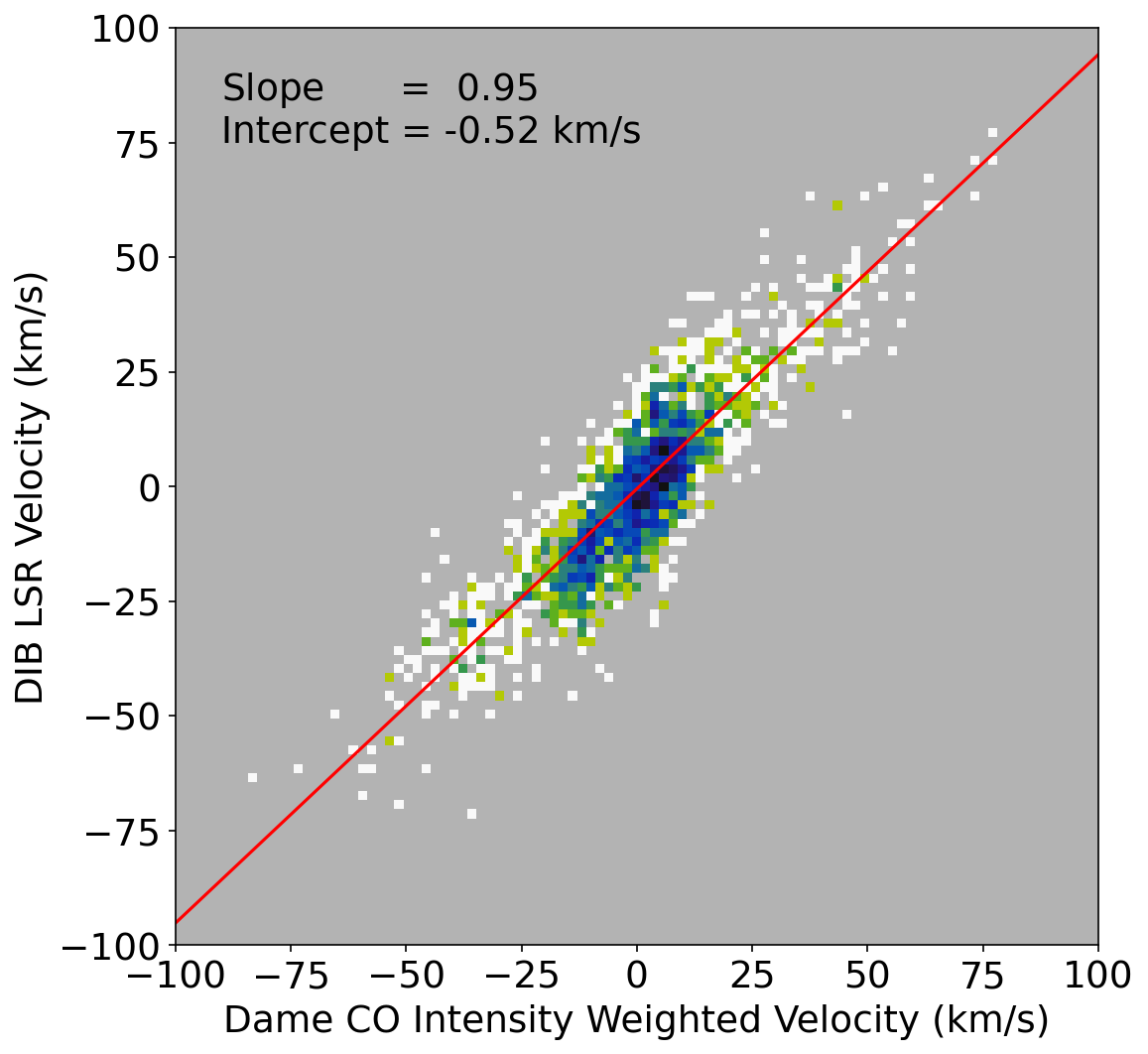}
\caption{Histogram (2D) illustrating correlation between the $v_{\rm{DIB}}^{\rm{LSR}}$ and intensity-weighted velocity of nearest CO component in the \citealt{Dame_2001_ApJ} CO map. The color scale indicates density, where darker colors indicate more detections and a logarithmic stretch has been applied.}
\label{fig:dust_co_corr}
\end{figure}

Other applications of the Galactic anti-center method to DIBs have conventionally performed a simple linear fit, holding $d$ to be a fixed constant, and reported the intercept as $\lambda_{\rm{rest}}$ and the slope in m\r{A}/deg. However, it is very difficult to compare these slopes between DIBs at different wavelengths or even the same DIB measured in samples with a different distribution of DIB carrier distances $d$ (see Equation \ref{eq:small_ang_gal}). Instead, our fit parameter in Equation \ref{eq:small_ang_gal} is $V_c = 204 \pm 60$ km/s. This can be easily compared to Galactic rotation models. This value is $<1$ $\sigma$ from that of \citealt{Bovy_2012_ApJ}, which is $216 \pm 6$ km/s. However, given the approximations made above, this should not be viewed as a measurement of $V_c$, but as a sanity check.

To compare to other work, we can compute the expected anti-center slope for a given $d$. Taking $d = 2$ kpc, we expect a slope of 19 m\r{A}/deg. Assuming a similar DIB carrier distance distribution, we can compare to \citealt{Zasowski_2015_ApJ} by multiplying by the ratio of the rest-frame wavelengths of the DIBs, $= 0.56$.\footnote{\citealt{Zasowski_2015_ApJ} restricted their fit to background stars within 2 kpc, which helps make this assumption more valid.} This would predict $32.1 \pm 4.5$ m\r{A}/deg. \citealt{Schultheis_2022_arXiv} report $45 \pm 20$ m\r{A}/deg, which uses the same DIB, but a different sample of detections and a clearly different $\lambda_{\rm{DIB}}$ (see Figure \ref{fig:contamfig}). However, the comparison is made more difficult by the fact that other work often only correct for the $U_{\rm{Sun}}$ velocity \citep{Zasowski_2015_ApJ,Schultheis_2022_arXiv}. The correction resulting from the $V_{\rm{Sun}}$ velocity is approximately linear with $\ell'$ near the Galactic anti-center and applying the correction reduces the resulting slope significantly.

With these validations (and those in Section \ref{sec:12CO}), we can combine $v_{\rm{DIB}}^{\rm{LSR}}$ with the Gaia location and parallax measurements to present a 4D map showing the radial velocity of DIB carriers as a function of position. Figure~\ref{fig:sky_vel} shows a projection of this map in angular ($\ell , b$) and Cartesian ($X, Y$) Galactic coordinates, taking an inverse variance weighted mean over the dimensions not shown. A lower noise map could be achieved with stricter DIB SNR cuts, though this comes at the expense of the number of detections and spatial coverage given the limited public Gaia RVS sample. The usual features (alternating sign quadrants) of a smooth Galactic rotation curve are observed. We defer searching for anomalies relative to a smooth Galactic rotation model until having a larger sample.

\vspace{3 mm}
\subsubsection{CO Comparison} \label{sec:12CO}

$^{12}$CO is a believed to be a good tracer of DIBs and dust because it has a well-known, easy-to-detect line and indicates the presence of carbonaceous molecules. We compare DIB kinematics to the \citealt{Dame_2001_ApJ} moment-masked, composite $^{12}$CO survey.\footnote{\href{https://lweb.cfa.harvard.edu/rtdc/CO/CompositeSurveys/}{https://lweb.cfa.harvard.edu/rtdc/CO/CompositeSurveys/}} We use the Full Galaxy cube ($-180\degree < \ell < 180\degree, |b| < 30\degree$) which is reported on a rectilinear Galactic Cartesian grid (1/8 $\degree$ spacing) and a velocity grid ($|v_{\rm{CO}}^{\rm{LSR}}| < 320$ km/s) with 1.3 km/s spacing. The LSR frame assumes the ``classic'' solar motion of 20 km/s toward (RA, Dec) $=$ (18h,30\degree) (epoch 1900), which is approximately $\vec{V}_{\rm{Sun}}$ = (10.3, 15.3, 7.7) km/s \citetext{private communication, Dame, 2022}. Figure~\ref{fig:VLSR_CO} shows the CO map, integrated along $b$.

In order to compare $v_{\rm{DIB}}^{\rm{LSR}}$ to the CO kinematics, we convert $v_{\rm{DIB}}^{\rm{LSR}}$ into the LSR frame convention used by \citealt{Dame_2001_ApJ} and over plot the DIB detections with DIB SNR $> 5.5$ in Figure \ref{fig:VLSR_CO}. The average $v_{\rm{DIB}}^{\rm{LSR}}$ as a function of $\ell$ follows the intensity-weighted average $v_{CO}$ (Figure \ref{fig:dust_co_corr}). Excitingly, we further observed a subpopulation of the DIBs that coincide with higher-amplitude oscillations in the CO map. Coloring the scatter plot by the distance to the background star, we find this subpopulation of DIB detections occur along lines of sight to more distant stars (compared to the average within the sample). This is consistent with the interpretation of higher-amplitude oscillations in $^{12}$CO as kinematically coherent Galactic substructure at larger distances, sometimes attributed to spiral arms.

We quantitatively demonstrate the correlation between $v_{\rm{DIB}}^{\rm{LSR}}$ in Figure \ref{fig:dust_co_corr}. Motivated by the multimodality shown in Figure \ref{fig:VLSR_CO}, before taking the intensity-weighted average of $v_{\rm{CO}}$, we find the peak in the CO spectrum nearest to $v_{\rm{DIB}}^{\rm{LSR}}$ for the ($\ell,b$)-pixel in the CO map corresponding to the line of sight for a DIB detection. Peak-finding has a tendency to slightly overstate the correlation, but the clumpiness of the CO necessitates such an approach. This peak-finding algorithm is discussed in detail in Appendix \ref{sec:dust_corr}, and we show there that a correlation similar to that shown in Figure \ref{fig:dust_co_corr} is obtained when restricting to low-velocity CO without any peak-finding. A linear least-squares fit, restricted to $|v_{CO}| < 25$ km/s, with inverse-variance weights using $\sigma(v_{\rm{DIB}}^{\rm{LSR}})$ is also shown in Figure \ref{fig:dust_co_corr}. The slope of $0.95 \pm 0.03$ illustrates a very strong 1:1 relationship between the $v_{\rm{DIB}}^{\rm{LSR}}$ and $v_{\rm{CO}}$, suggesting that where they coexist, DIB carriers and CO are predominately co-moving.

The intercept of the linear correlation in Figure \ref{fig:dust_co_corr} provides a semi-independent check on the $\lambda_{\rm{rest}}$ and its associated uncertainty reported in Section \ref{sec:Kin} because the frequency of the CO transition is known to high precision. An offset of $-0.52 \pm 0.25$ km/s corresponds to a $\Delta\lambda_{\rm{rest}} = -0.015$ \r{A}, or $< 0.2 \sigma$ given the reported $\lambda_{\rm{rest}}$ uncertainty. Further, the scatter of residuals/$\sigma(v_{\rm{DIB}}^{\rm{LSR}})$, or Z-scores, was 0.6, indicating that the reported uncertainties slightly overestimated the scatter in the residuals. However, given the selection bias toward correlation given the use of a peak-finding algorithm, a suppression of the scatter in the residuals is to be expected.

Thus, correlation with $v_{\rm{CO}}$ provides a secondary semi-independent validation of $\lambda_{\rm{rest}}$, $v_{\rm{DIB}}^{\rm{LSR}}$, and their associated uncertainties, which lends confidence to the use of the MADGICS DIB catalog for kinematics studies.

\subsection{Moment-2} \label{sec:Mom2}

The width of an absorption feature (moment-2) can often be used to infer properties about the physical environment or transition itself. In radio observations of the ISM, line widths are often attributed to a combination of Doppler, supersonic turbulence, and opacity broadening \citep{Hacar_2016_AA}. However, given the wavelength and observed line width for the DIBs, simple order of magnitude calculations show that these mechanisms are unlikely to contribute significantly to the line width \citep{Edwards_1994_AIPC}. The dominant intrinsic source of broadening is lifetime broadening, which agrees with the very short excited state lifetimes predicted for most DIB carrier candidates \citep{Snow_2002_ApJ,Campbell_2015_Natur}.

\begin{figure}[t]
\centering
\includegraphics[width=\linewidth]{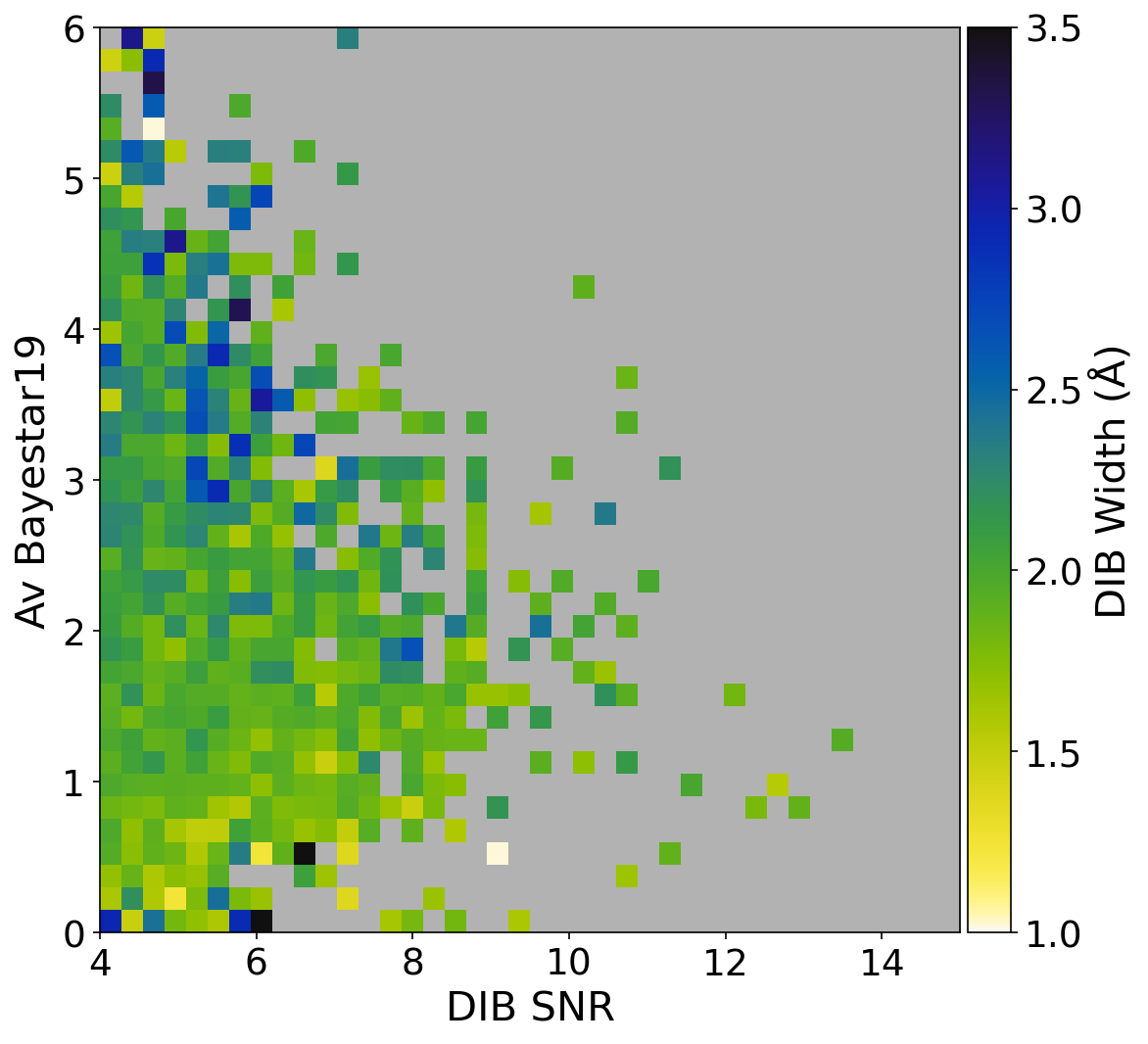}
\caption{Median of the DIB width $\sigma_{\rm{DIB}}$ as a function of the DIB SNR and the extinction along the line of sight in the Bayestar19 3D dust map \citep{Green:2019:ApJ:}. The weak correlation could be due to the presence of multiple cloud components at different velocities along the line of sight and/or the presence of velocity structure within individual clouds.}
\label{fig:AvSigma}
\end{figure}

The observed line width can also appear broader because of heterogeneity in the central wavelength. For example, multiple dust clouds moving at different velocities along the line of sight can lead to multiple unresolved peaks \citep{Tchernyshyov_2018_AJ}. As an order of magnitude, a difference in velocity of 20 km/s would lead to a wavelength shift that is only half of the typical $\sigma_{\rm{DIB}}$, making resolving two components in the spectra extremely difficult. We expect this to be one of the largest sources of the variability in observed $\sigma_{\rm{DIB}}$. Heterogeneity in the chemical environment, which can slightly shift the transition wavelength for some carriers in the column density probed by the line of sight, is another possible source of broadening. However, given that most DIB carriers remain unidentified, it is difficult to make predictive theories about the role of chemical environment heterogeneity.

We observe (Figure \ref{fig:AvSigma}) a weak positive correlation of $\sigma_{\rm{DIB}}$ with the extinction (Av) from Bayestar19 \citep{Green:2019:ApJ:}. This could result from there being more components (clouds at different velocities) on average along the lines of sight with higher reddening. It could also result from different environments in higher extinction clouds. Because many of the highest reddening lines of sight are also low signal-to-noise and we show a positive detection bias in $\sigma_{\rm{DIB}}$, we hold off on rigorous interpretation of $\sigma_{\rm{DIB}}$ until a larger sample of Gaia RVS spectra are available so as to have more statistical power.


\section{DIB Model} \label{sec:Model}

We provide a brief introduction to MADGICS and the specific choices made in the application to Gaia RVS spectra, with a more detailed description of the method in future work.

\subsection{Statistical Description} \label{sec:Stats}

MADGICS is a component separation algorithm which decomposes an observation, a data vector $x_{\rm{tot}}$, into a linear combination of components $x_k$. In order to speak about such a decomposition, one must have prior information about the different components, which we require to sum to the data, and about the properties of each component. In MADGICS, we express that prior as a pixel-pixel covariance matrix $C_k$ in the data space. In this case, this is literally the covariance of wavelength bin $i$ and wavelength bin $j$ in the spectrum (Equation \ref{eq:decomp_set}). Prior information about the mean of a component $\mu_k$ can also be included, though in practice, we often set $\mu_k = 0$. Here, we have implicitly assumed there is no cross-component covariance in the priors. For notational convenience, we define the sum of the covariance and mean priors over all components (Equation \ref{eq:tot_def}).

\begin{ceqn}
\begin{align} \label{eq:decomp_set}
    x_{\rm{tot}} &= \sum_{k=1}^{N_{\rm{comp}}} x_k \quad \text{where} \quad x_k \sim \mathcal{N}(\mu_k,C_k) \\
    \label{eq:tot_def}
    C_{\rm{tot}} &= \sum_{k=1}^{N_{\rm{comp}}} C_k \quad \quad \mu_{\rm{tot}} = \sum_{k=1}^{N_{\rm{comp}}} \mu_k
\end{align}
\end{ceqn}

To obtain a posterior for the contribution of each component to the signal, we specify a likelihood that is the product of the Gaussian probability for $\hat{x}_k$ given $\mathcal{N}(\mu_k,C_k)$ for each of the components and then impose the constraint that the components sum to the data to obtain a posterior. The posterior mean and covariance per component can be obtained analytically as
\begin{ceqn}
\begin{align} \label{eq:post_mean}
    \hat{x}_{k} &= C_k C_{\rm{tot}}^{-1}\left(x_{\rm{tot}} -\mu_{\rm{tot}}\right) + \mu_k \\
    \label{eq:post_kk}
    \hat{C}_{kk} &= (I - C_k C_{\rm{tot}}^{-1})C_k \\
    \label{eq:post_km}
    \hat{C}_{km} &= -C_k C_{\rm{tot}}^{-1} C_m
\end{align}
\end{ceqn}
where $I$ is the identity matrix, $\hat{x}_k$ is the predicted mean component, $\hat{C}_{kk}$ is the predicted pixel-pixel covariance of pixels in component $k$, and $\hat{C}_{km}$ is the predicted covariance of a pixel in component $k$ with a pixel in component $m$. \aks{See Appendix \ref{sec:pixCov} for an example of these predicted (posteriors) covariances.}

In terms of computational ease, Equations \ref{eq:post_mean}--\ref{eq:post_km} show that the solution depends on only a single matrix inverse $C_{\rm{tot}}^{-1}$, which can be computed quickly using low-rank approximations (see Section \ref{sec:Prior}) and Woodbury updates \citep{max1950inverting}. In this work, processing one million Gaia RVS spectra took only 650 core-hours.\footnote{Production runs were executed on AMD 64-core EPYC 7702P processors (AVX2) with 480 GB RAM running a Linux kernel v4.18.0 (Rocky Linux).}

In MADGICS, all of the difficulty lies in specifying the models and priors on each component in the model. For Gaia RVS spectra, we perform two decompositions modeling the data as either star + residual or star + DIB + residual. The sum of $\hat{x}_{\rm{DIB}}$ and $\hat{C}_{\rm{DIB},\rm{DIB}}$ give the estimates of $EW_{\rm{DIB}}$ and the uncertainty in $EW_{\rm{DIB}}$, respectively.

\vspace{3mm}
\subsection{Prior Construction} \label{sec:Prior}

\subsubsection{Stellar Prior} \label{sec:stellarprior}
The stellar covariance is built from a high ($>70$) stellar SNR and low reddening (SFD $E(B-V) < 0.05$ mag) training set, \aks{where we have used a cut on a 2D infrared emission-based dust map that acts an upper bound on the dust along the line of sight (\cite{Schlegel_1998_ApJ}, SFD). W}e also require GSP-Spec Flags 1-13 be zero (the highest quality value) in order to be most exclusive against defects and outliers. After cuts, this is 39,657 spectra. For simplicity, we slightly restricted the spectral range of the stellar model to 8469.5 - 8688.3 \r{A} (2189 versus 2401 pixels) where the vast majority of all Gaia RVS spectra had coverage so that we can work in a single common data space.

\begin{figure}[t]
\centering
\includegraphics[width=\linewidth]{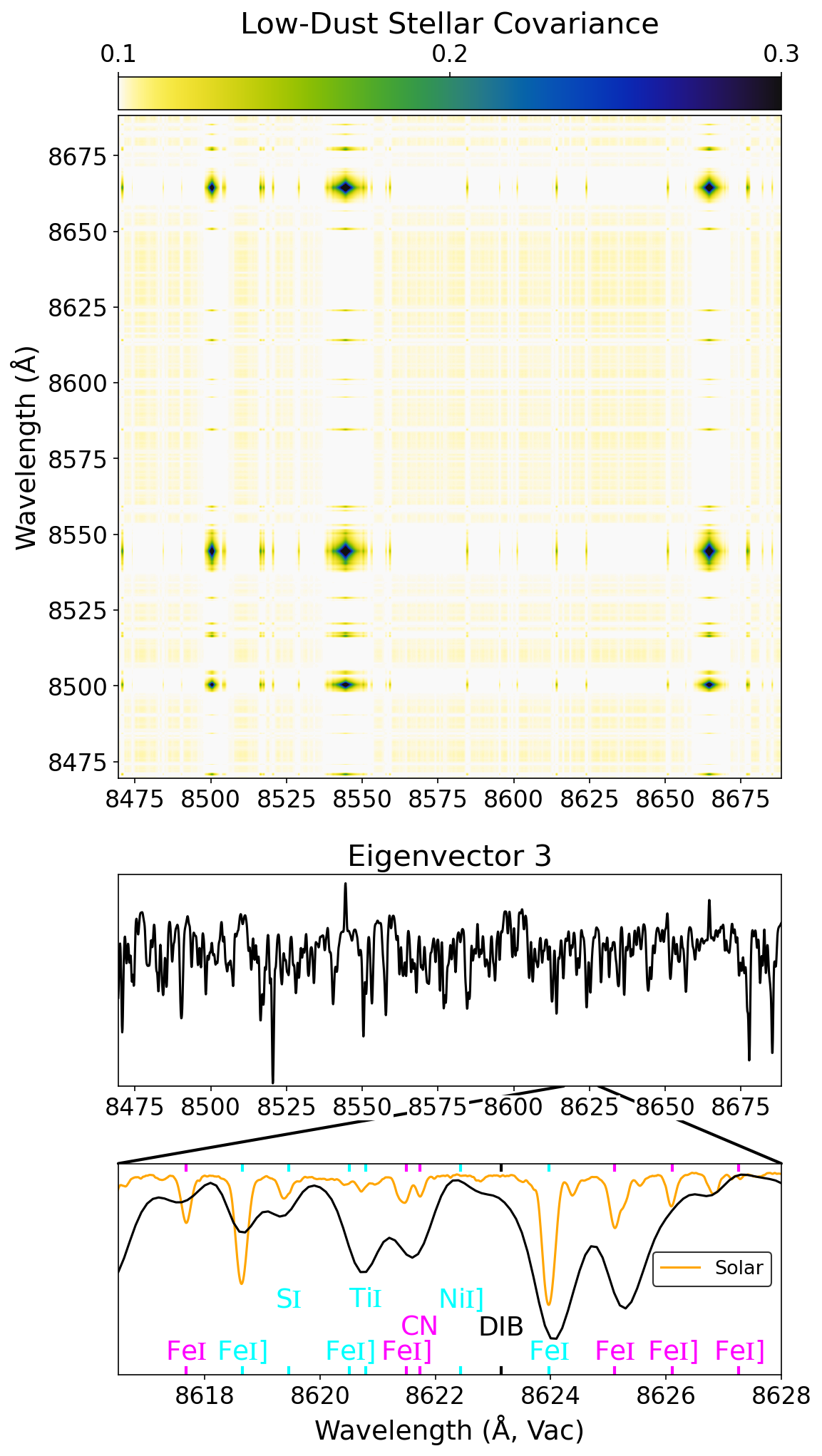}
\caption{(Top) Covariance matrix of high stellar SNR and low-dust spectra used as MADGICS prior for stellar component. (Middle) Third eigenvector of stellar covariance matrix. (Bottom) Zoom-in of third eigenvector near the DIB feature. Guidelines for the central wavelength of the DIB and stellar lines are labeled and color coded. A higher-resolution solar spectrum from \citealt{1984sfat.book.....K} is plotted in orange for reference.
}
\label{fig:stellar_cov}
\end{figure}

\begin{figure*}[t]
\centering
\includegraphics[width=\linewidth]{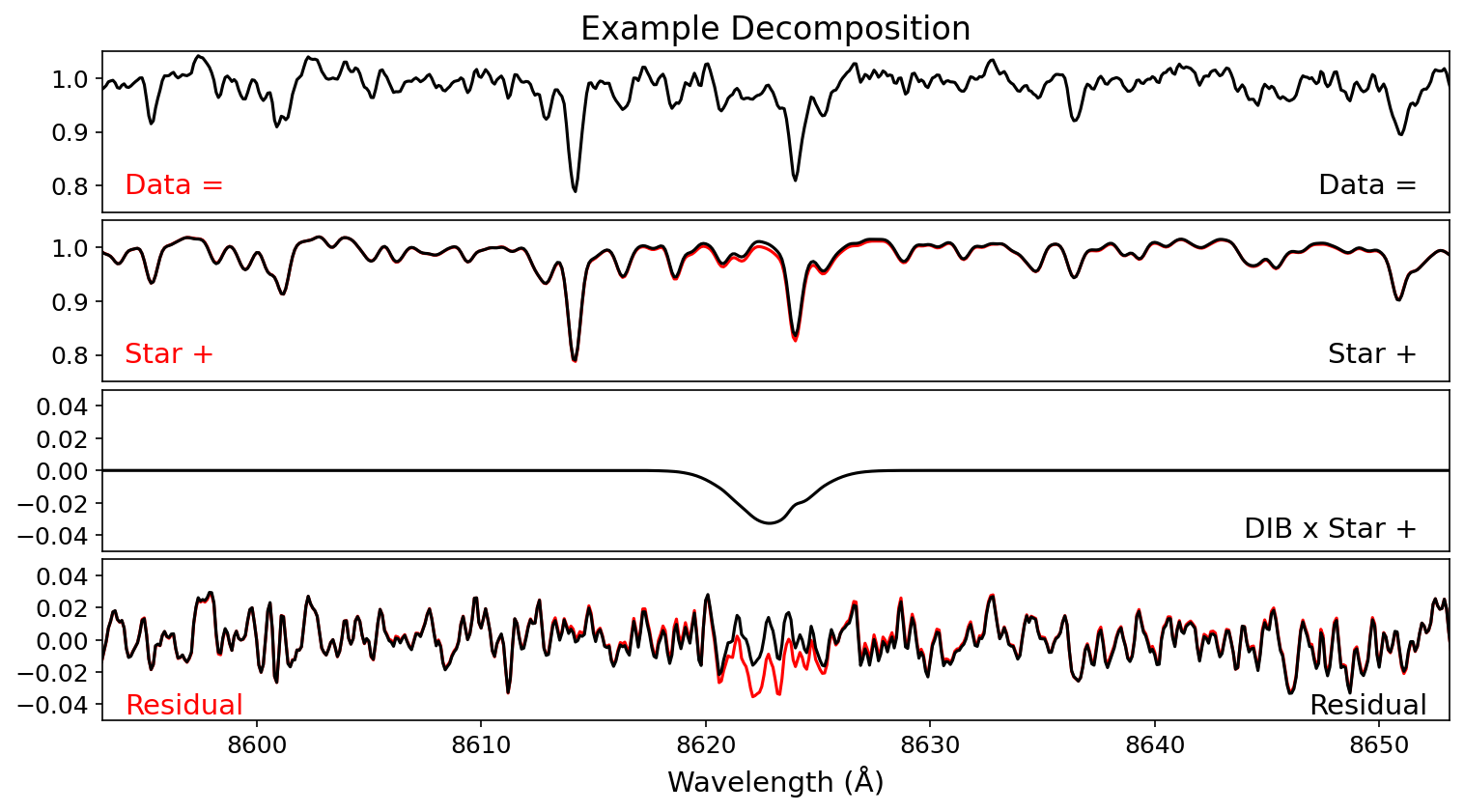}
\caption{Example MADGICS decomposition of Gaia RVS spectrum under a star + residual (red) and star + DIB $\times$ star + residual model (black).
}
\label{fig:decomp}
\end{figure*}

Since the Gaia RVS spectra are continuum normalized, we subtract off an average constant continuum of $0.95$. In the notation of Section \ref{sec:Stats}, this is letting $\mu_{\rm{star}} = 0.95$ times the ones vector. The covariance is then built by multiplying the matrix of these observations by its transpose, with the relative contributions weighted by stellar SNR.\footnote{Weights were capped at 300 to prevent any one spectrum from dominating the covariance.} However, since the continuum normalization is not stable (see Appendix \ref{sec:selectFun}), we add $0.1$ times the ones matrix to the covariance so that the stellar component will absorb overall shifts in the continuum normalization.\footnote{The value 0.1 was chosen to be comparable to average pixel in the covariance matrix far from the diagonal and clear stellar features. \aks{However, the component separation is extremely stable to changes in the choice of this value. Similar results are also obtained by manually adding an eigenvector of ones to the low-rank approximation (described below) instead of adding a constant times the ones matrix to the full covariance matrix (as is described here).}}

The resulting stellar covariance matrix (the MADGICS prior) is shown in Figure \ref{fig:stellar_cov}. Despite the complexity of stellar spectra, this covariance is in general well-described by a low-rank approximation, which also significantly accelerates the computation. We keep the first 50 eigenvectors to describe the stellar covariance, one of which is shown in Figure \ref{fig:stellar_cov}.\footnote{\aks{The number of eigenvectors in the low-rank approximation was chosen in part by inspecting the steep fall-off of the eigenspectrum. We also ran empirical tests that showed negligible changes in the stellar component obtained during the decomposition of test spectra as a function of the dimension of the low-rank approximation by 50.}} We explicitly check for and do not find significant peaks centered at the DIB wavelength.

\begin{figure*}[t]
\centering
\includegraphics[width=\linewidth]{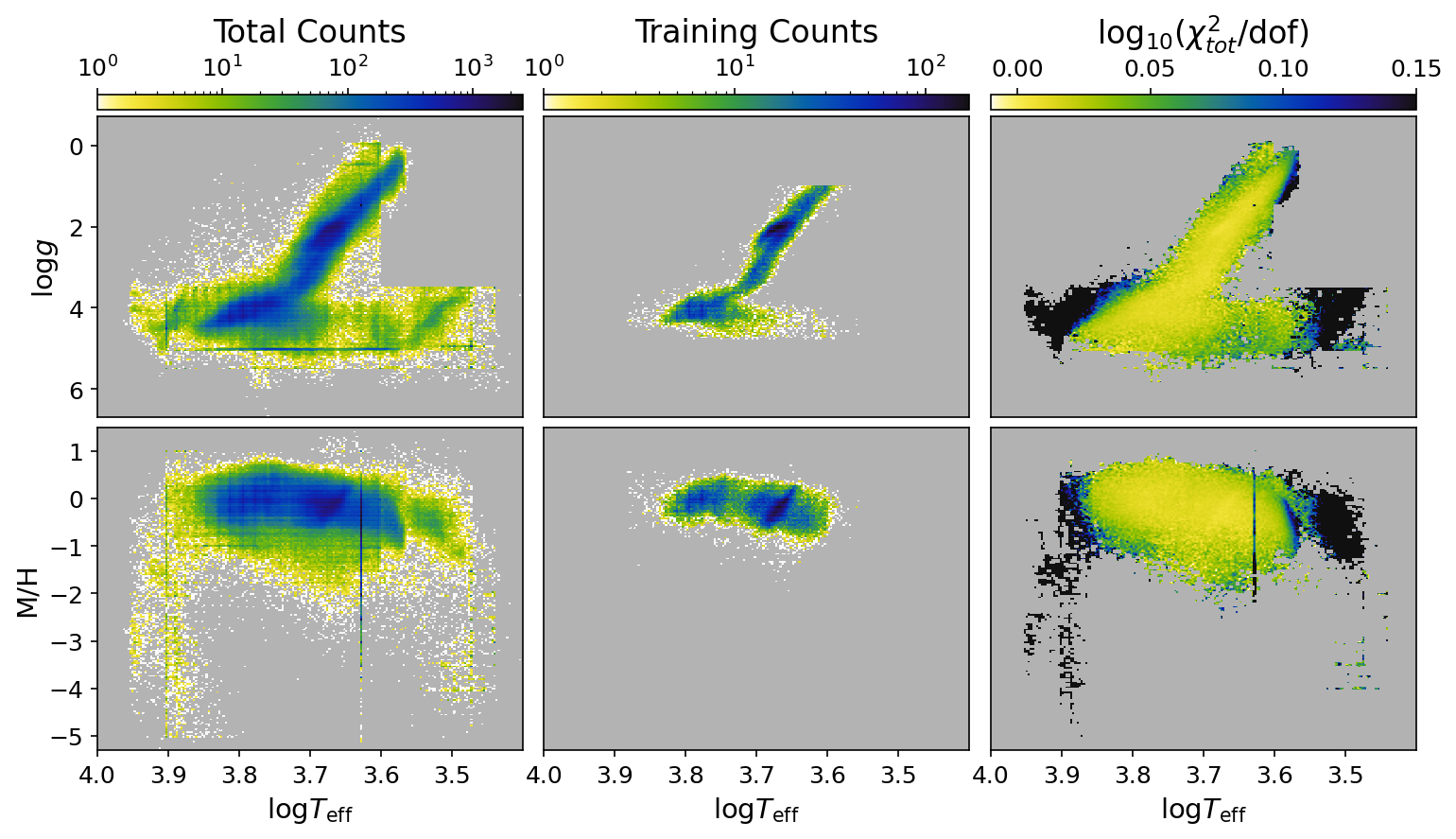}
\caption{Histograms (2D) in the space of GSP-Spec stellar parameters for all (left) and training (middle) spectra. (Right) The overall ``goodness of fit'', measured by how close $\chi^2$/dof is to 1, is shown. Each pixel is the median of $\log_{10}(\chi^2$/dof), which is zero in the ideal case.
}
\label{fig:stellar_space}
\end{figure*}

In Figure \ref{fig:stellar_cov}, we overplot the same guidelines as in Figure \ref{fig:contamfig}. We show that there are significant peaks, comparable to those for lines carefully calibrated in the GSP-Spec line list, centered at the wavelengths of the pile-ups identified in Figure \ref{fig:contamfig} as having the largest impact on biasing the Gaia DIB catalog. By plotting a higher resolution solar spectrum from \citealt{1984sfat.book.....K} for reference, we see that the features in our stellar model that are coincident with those pile-ups also coincide with large amplitude features in the solar spectrum. Here we adopt the largest amplitude lines from the Kurucz archive (\href{http://kurucz.harvard.edu/}{http://kurucz.harvard.edu/}) contributing to those features in high resolution spectra of both the Sun \citep{1984sfat.book.....K} and Arcturus \citep{2000vnia.book.....H} as labels for the respective pile-up locations (see Appendix \ref{sec:linelist}). However, this comparison confirms that the contamination in the Gaia DIB catalog shown in Figure \ref{fig:contamfig} is of stellar origin, regardless of the specific line assignments of those features.

\subsubsection{Residual Prior} \label{sec:residprior}

In all measurements, there is contribution from noise that will follow its own covariance structure, and thus we also include a ``residual'' component to capture this noise. Ideally, these measurement uncertainties are uncorrelated and the residual prior is a simple diagonal. In that case, unmodeled features in the data which do not conform well to the prior of any other component in the model  (such as a previously unidentified DIB) will simply appear in the residuals.

Gaia DR3 includes a per-pixel flux uncertainty vector, which could be used on the diagonal of the residual covariance prior. However, when doing so, one finds that the MADGICS residual component in fact has correlations, with a kernel that appears to be the result of upstream processing interpolating the spectra onto a common grid.\footnote{The kernel has the form $[0.1, 0.4, 0.8, 1.0, 0.8, 0.4, 0.1]$ centered on a given pixel.} These correlations impact the uncertainty of $EW_{\rm{DIB}}$, but we want to retain a diagonal residual prior, so we inflate the residual prior by multiplying by a constant ($3.6$) such that the sum of the correlated noise and uncorrelated noise covariances are the same.

\subsubsection{DIB Prior} \label{sec:DIBprior}

The contribution from DIBs is fundamentally multiplicative since it is an absorption of the background starlight. Since MADGICS is additive, it finds a linear decomposition of the data into components, we first model the spectra (even one containing a DIB detection) as being star + residual components (Figure \ref{fig:decomp}). Then, we express a prior on the line shape of the DIB absorption. In this case, we use a Gaussian for simplicity, but since we need only specify the DIB prior as a pixel-pixel covariance matrix, we can accommodate arbitrary line shapes and/or degrees of asymmetry. Sampling over variations in the line shape, we construct a pixel-pixel covariance matrix. 

Here we keep only a two-component, low-rank approximation for the DIB covariance, which we generate analytically using the functional form of a Gaussian and its derivative with respect to wavelength. We use a constant amplitude prior of 0.1 and reduce the amplitude of the derivative by a factor of 0.01. \aks{The amplitude value was chosen to be large relative to the expected DIB signal (a ``loose'' prior). MADGICS is able to use such ``loose'' priors and be insensitive to the specific value chosen because it leverages the correlation of many pixels. Stated another way, low-dimensional slices in high-dimensional spaces are fairly orthogonal. The derivative value was chosen to marginalize over shifts in the observed wavelength smaller than the step size of the radial velocity scan (see below).}

However, this covariance is a prior on $x'_{\rm{DIB}}$ while the linear contribution of the DIB to the spectrum is $x_{\rm{DIB}} = x'_{\rm{DIB}} \times x_{\rm{star}}$. Since we do not know $x_{\rm{star}}$, we use $\hat{x}_{\rm{star}}$ from the star + residual decomposition and rescale the covariance for $x'_{\rm{DIB}}$ to obtain the covariance for $x_{\rm{DIB}}$. Since the contribution from the DIB is predominantly in the residual component in the star + residual model, the $\hat{x}_{\rm{star}}$ changes only slightly in the star + DIB + residual model (see Figure \ref{fig:decomp}). We can then update $\hat{x}_{\rm{star}}$, using the component from the star + DIB + residual model, and iterate until convergence. In practice, we only iterate to refine the $x_{\rm{DIB}}$ once.

\begin{figure}[b]
\centering
\includegraphics[width=\linewidth]{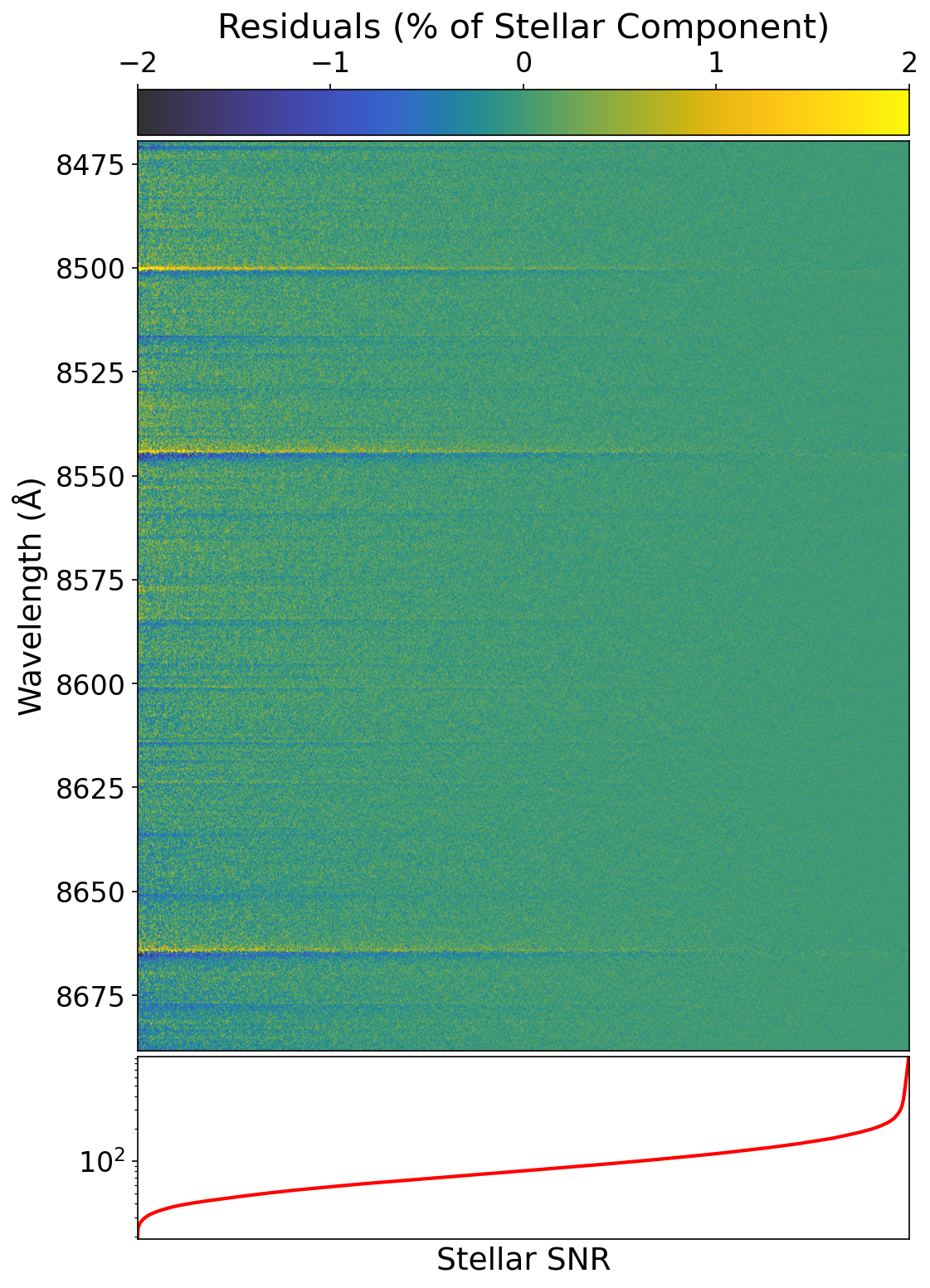}
\caption{Residual components, sorted by stellar SNR, divided by the stellar component, and averaged over 32 adjacent spectra. Bottom panel shows value of stellar SNR along the sorted axis.
}
\label{fig:residuals}
\end{figure}

While MADGICS can easily marginalize over (co)-variations in peak amplitudes, it struggles to marginalize over shifts in observed wavelength (radial velocity) or changes in peak width ($\sigma_{\rm{DIB}}$) which are not well described in the data-space representation. In order to optimize over radial velocity and $\sigma_{\rm{DIB}}$, we build a grid of covariance priors over a range of ($\lambda_{\rm{DIB}}$, $\sigma_{\rm{DIB}}$) and evaluate the $\Delta\chi^2$ at each point, seeking the minimum (most negative) $\Delta\chi^2$. 

In practice, we find the search of the 2D ($\lambda_{\rm{DIB}}$, $\sigma_{\rm{DIB}}$)-space well-approximated using line searches with a single iteration of refinement. The $\lambda_{\rm{DIB}}$ scan is $\pm 100$ pixels, in 0.1 pixel steps (0.01 \r{A}) at highest resolution. The $\sigma_{\rm{DIB}}$ scan is 0.4 to 4 \r{A}, in 0.01 \r{A} steps at highest resolution (see code in Section \ref{sec:dataavil} for more). The resolution of the $\lambda_{\rm{DIB}}$ and $\sigma_{\rm{DIB}}$ scans were chosen such that the step size is $\leq10\%$ the typical uncertainty in each parameter estimate.

The $\lambda_{\rm{DIB}}$ and $\sigma_{\rm{DIB}}$ uncertainties are estimated from the shape of the $\Delta\chi^2$ surface around the minimum, using the Hessian estimated with finite differences. For a given point in the ($\lambda_{\rm{DIB}}$, $\sigma_{\rm{DIB}}$)-grid, we have estimates of $EW_{\rm{DIB}}$ and the uncertainty in $EW_{\rm{DIB}}$ given by sum of $\hat{x}_{\rm{DIB}}$ and $\hat{C}_{\rm{DIB},\rm{DIB}}$, respectively. However, in order to marginalize over $\sigma_{\rm{DIB}}$, we combine these Gaussian estimates of the $EW_{\rm{DIB}}$ for several $\sigma_{\rm{DIB}}$ near the minimum. These estimates are combined with relative weights given by their likelihood ratios.

The final optimal decomposition for the example in Figure \ref{fig:decomp} is shown in black. The substructure in the DIB $\times$ star component is the result of the stellar features, which is removed upon dividing by the stellar component. The DIB $\times$ star and residual components are on the same scale and have comparable amplitudes. Note that the wavelength range in Figure \ref{fig:decomp} is intermediate to the full RVS spectral range and the window near the DIB shown in Figure \ref{fig:contamfig} and \ref{fig:stellar_cov}. The DIB profile is fairly wide relative to the narrow window shown previously when focusing on stellar lines.

\subsection{Stellar Validation} \label{sec:EmpValid}

While we do not focus on modeling the stars, we briefly illustrate the quality of stellar modeling because it impacts which background stars can be used to detect DIBs and the magnitude of any possible biases resulting from stellar features. In Figure \ref{fig:stellar_space}, we show both the distribution of the stellar training set and all public Gaia DR3 RVS spectra as a function of $\log g$, \aks{[M/H]}, and $\log T_{\rm{eff}}$ as estimated by GSP-Spec. The training set is an order of magnitude smaller, but has a similar shape compared to the full distribution. Only populations at the extremes in temperature and metallicity are notably missing, populations which form a small fraction of the overall sample. 

Measuring how well spectra outside the training set are modeled (generalization) is one way to understand the flexibility and validity of the stellar prior. We do this by showing the median $\log(\chi^2_{\rm{tot}}$/dof) for spectra as a function of the stellar parameters. The color scale is saturated at the value used as a ``goodness of fit'' cut in making the catalog (see Table \ref{tab:dibcat} and Section \ref{sec:Comp2Cat}). This highlights regions of stellar parameter space that are so poorly modeled as to be excluded from consideration in making the DIB catalog. As shown in Figure \ref{fig:stellar_space}, only a very small fraction of the sample at the most extreme stellar parameters are excluded. 

Over the majority of the stellar parameter space, the $\chi^2_{\rm{tot}}$/dof is uniform and near one. This stable region is notably much larger than that spanned by the training set, indicating good model generalization. We also checked for any trends of the inferred DIB parameters as a function of stellar parameters and found no clear correlations. However, the occasional negative $EW_{\rm{DIB}}$ detections do preferentially occur in relatively less common stellar types (in lower density regions of the left two columns of Figure \ref{fig:stellar_space}).

We also examine the residuals for evidence of systematic mismodeling of stellar features. We sort as a function of stellar SNR and average over blocks of 32 spectra to suppress random noise and emphasize features of stellar origin (Figure \ref{fig:residuals}). Clear features associated with the Ca triplet (8500.36, 8544.44, 8664.52\,\r{A}) appear and increase in magnitude with decreasing stellar SNR, reaching $\sim2$\% of the stellar component at the lowest stellar SNR. 

\aks{The Ca triplet residuals resemble the derivative of the line shape and are indicative of a shift in the reported rest frame of the spectra (a shift in the apparent wavelength of the Ca triplet). As shown in Appendix \ref{sec:velShift}, this shift is $\sim 300$ m/s at low stellar SNR. The detection of this shift both demonstrates the extraordinary sensitivity of MADGICS for estimating radial velocity measures and is entirely consistent with the reported precision of Gaia RVS.} Other stellar features are visible to a lesser extent, $<1\%$ in magnitude. Since synthetic models have residuals $\sim5\%$ for the Ca triplet \citep{AllendePrieto_2013_A_A}, the residuals shown in Figure \ref{fig:residuals} indicate high-quality modeling of the stellar component.


\section{Catalog Building} \label{sec:Build}

\subsection{Components to Catalog} \label{sec:Comp2Cat}

\begin{figure}[t]
\centering
\includegraphics[width=\linewidth]{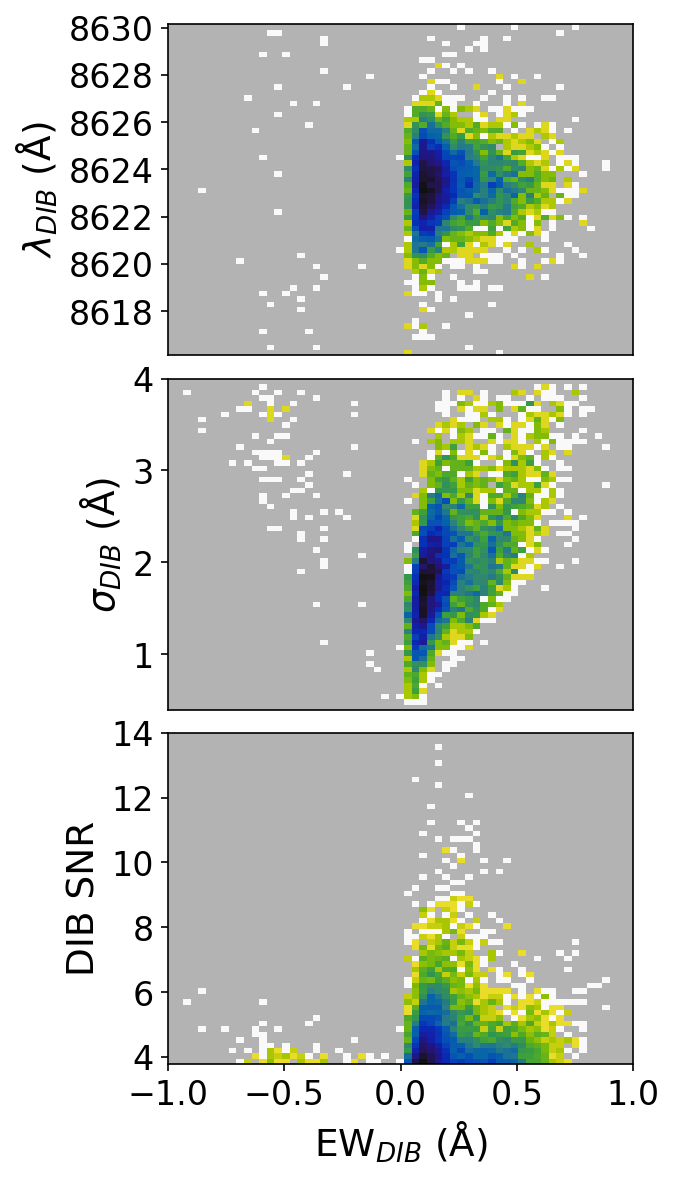}
\caption{Histograms (2D) showing projections of the MADGICS catalog in the space of moment-0, moment-1, moment-2, and DIB SNR. The color scale indicates density, where darker colors indicate more detections, and each panel has its own logarithmic stretch.
}
\label{fig:cat_dist}
\end{figure}

\begin{figure*}[t]
\centering
\includegraphics[width=\linewidth]{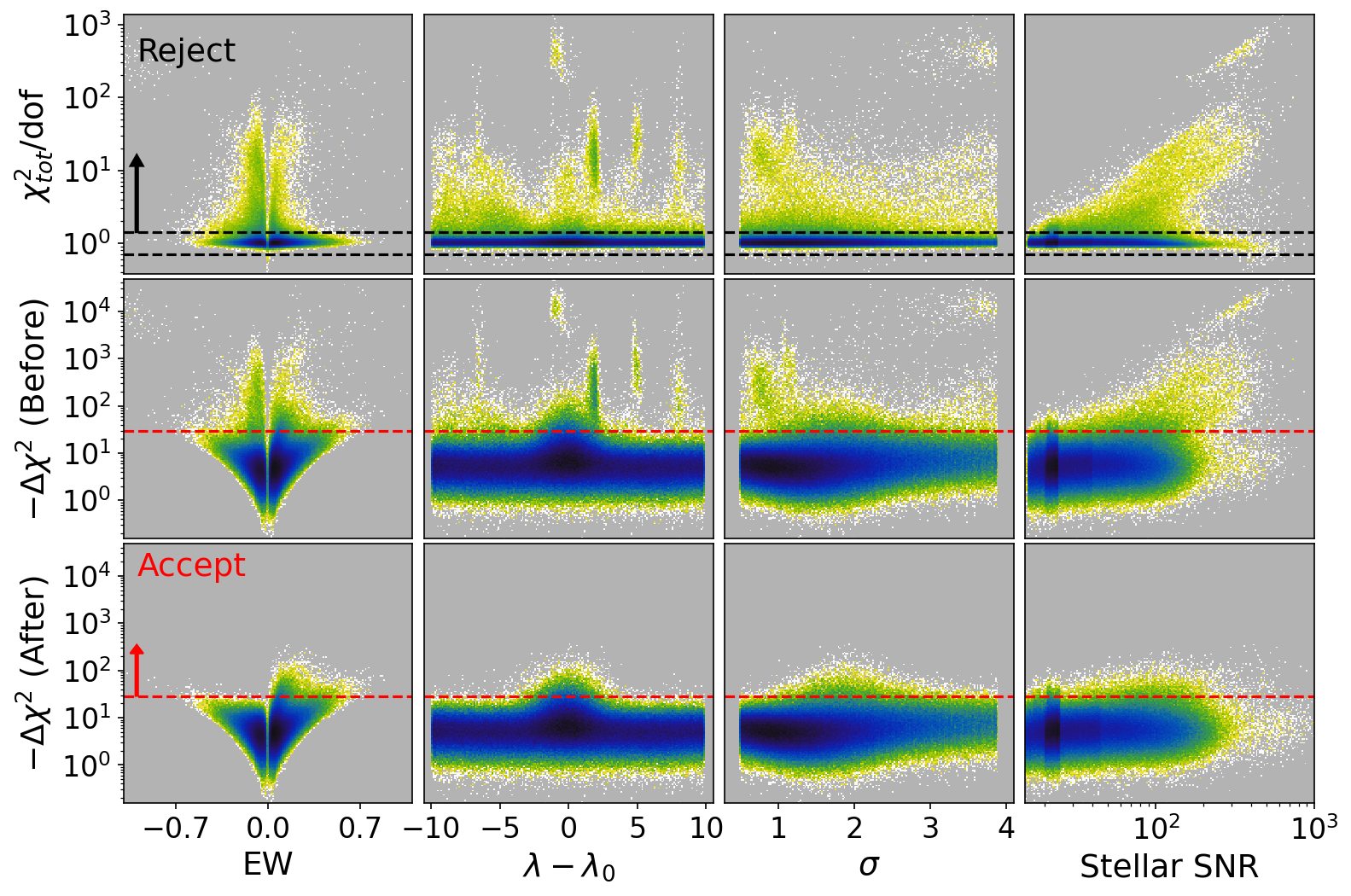}
\caption{Histograms (2D) for DIB models as a function of $EW_{\rm{DIB}}$, $\lambda_{\rm{DIB}}$, $\sigma_{\rm{DIB}}$, and stellar SNR in the columns, left to right respectively. The $\lambda_{\rm{DIB}}$ is reported relative to the reference center of the MADGICS parameter grid. Positive $EW_{\rm{DIB}}$ indicates absorption. The rows show this distribution for (top) the total $\chi^2$ per degree of freedom of the component fit, the negative change in $\chi^2$ as a result of adding the DIB component before (middle) and after (bottom) imposing cuts on $\chi^2_{\rm{tot}}/\rm{dof}$. Guidelines indicate cuts imposed in each space. The color scale indicates density, where darker colors indicate more detections, and each panel has its own logarithmic stretch.}
\label{fig:chi2_tot}
\end{figure*}

Below we describe how to move from the MADGICS components to a DIB detection catalog (Figure \ref{fig:cat_dist}). In obtaining the MADGICS components for all public spectra with the model described in Section \ref{sec:Model}, we implicitly drop spectra without full coverage in the slightly restricted spectral range of the stellar model (8469.5 - 8688.3 \r{A}). These are predominantly stars with extraordinarily high radial velocities or defects in the mean RVS spectra; this cut removes only 754/999,645 spectra. We also cut DIB ``detections'' which occur too close to the boundary of the $\lambda_{\rm{DIB}}$ and $\sigma_{\rm{DIB}}$ parameter grids, such that the uncertainty of $\lambda_{\rm{DIB}}$ and $\sigma_{\rm{DIB}}$ cannot be estimated via a Hessian computed using finite differences. We also remove a small number of cases where the Hessian has inverted curvature along either the $\lambda_{\rm{DIB}}$ or $\sigma_{\rm{DIB}}$ direction for the step size chosen. These are detections which do not have robust minima.\footnote{Most of the above cases would be removed by the $\chi^2_{\rm{tot}}$ cut below, but we remove them first for clarity in Figure \ref{fig:chi2_tot}.} 

Since the MADGICS DIB component need not be exactly Gaussian, we extract the Gaussian moments of the component through a simple least-squares fit of a Gaussian function. These $\lambda_{\rm{DIB}}$ and $\sigma_{\rm{DIB}}$ estimates agree quite well with the MADGICS estimates after the DIB SNR and $\chi^2_{\rm{tot}}$ cuts have been applied and there are no identifiable systematics. Before cuts, there are outliers associated with poorly modelled stellar spectra or cases where $EW_{\rm{DIB}}$ is in the noise. For that reason, we also cut ``detections'' where the least-squares fit $\lambda_{\rm{DIB}}$ and $\sigma_{\rm{DIB}}$ are at the edge the MADGICS search grid. After edge-of-grid cuts, 620,853 spectra remain.

We first apply the cut on $0.71 < \chi^2_{\rm{tot}}/\rm{dof} < 1.41$.\footnote{Throughout, we have rescaled the $\chi^2_{\rm{tot}}$ by the factor of $3.6$ mentioned in Section \ref{sec:residprior} to obtain the usual interpretation of the $\chi^2_{\rm{tot}}/\rm{dof}$.} The values for this cut were chosen to be a conservative definition of a ``reasonably good'' fit and were validated on injection tests (see Section \ref{sec:Inject}). As shown in Figure \ref{fig:chi2_tot}, this cut can clearly separate errant detections that are locked to a specific wavelength in the stellar frame. This outlier population in $\chi^2_{\rm{tot}}/\rm{dof}$ also has unusually small or large $\sigma_{\rm{DIB}}$. Clear outlier populations with respect to stellar SNR are also removed, the origin of which we have not specifically investigated, but appear related to the spectral normalization effects discussed in Appendix \ref{sec:DIBSNR}. Note that all of these detections have a large corresponding $\Delta\chi^2$ and so would otherwise be easily mistaken as significant detections of DIBs. However, while the addition of the DIB component significantly improves the modeling of these spectra, these cases can be easily removed by recognizing that the overall goodness of fit remains poor. After the $\chi^2_{\rm{tot}}/\rm{dof}$ cut, 586,872 spectra remain.

\begin{figure*}[t]
\centering
\includegraphics[width=\linewidth]{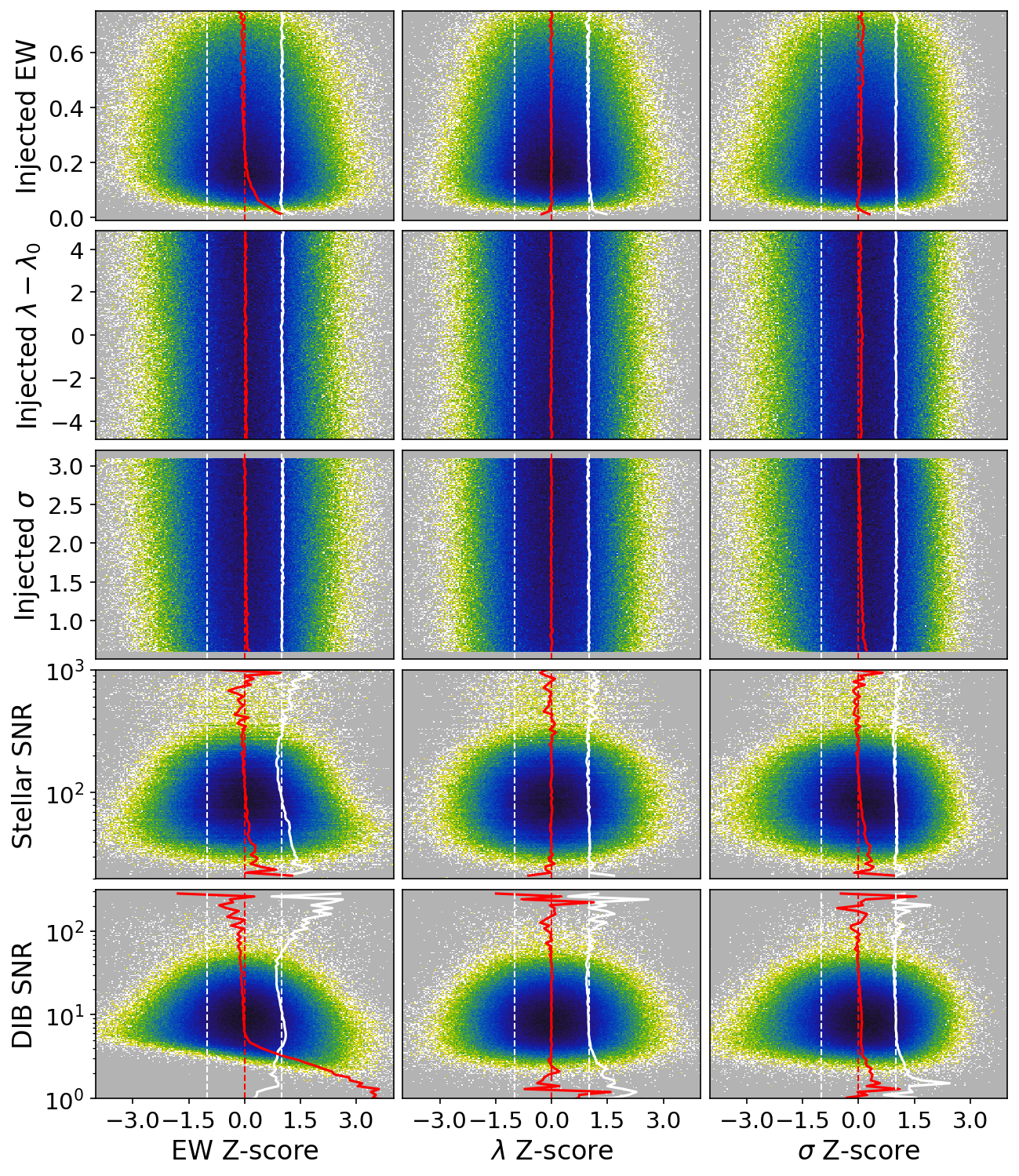}
\caption{Histograms illustrating the bias of DIB detections relative to reported uncertainties (Z-scores) for injection tests where ground truth is known. Histograms are a function of injected $EW_{\rm{DIB}}$, injected $\lambda_{\rm{DIB}}$, injected $\sigma_{\rm{DIB}}$, stellar SNR and DIB SNR in the rows, and show the bias in the recovered quantities $EW_{\rm{DIB}}$, $\lambda_{\rm{DIB}}$, $\sigma_{\rm{DIB}}$ in the columns (left to right). The $\lambda_{\rm{DIB}}$ is reported relative to the reference center of the MADGICS parameter grid. The color scale indicates density, where darker colors indicate more detections, and each panel has its own logarithmic stretch. Dashed guidelines provide reference for $0$ and $\pm 1 \sigma$. Solid lines indicate the center (red) and spread (white) of the Z-score distribution.
}
\label{fig:inject}
\end{figure*}

After these spuriously large $\Delta\chi^2$ cases have been removed, we make the final cut on $\Delta\chi^2$ such that the DIB SNR $> 3.8$, where\footnote{This $\Delta\chi^2$ does not include the likelihood volume term associated with the DIB component.}
\begin{ceqn}
\begin{align}
    \rm{DIB \ SNR} &= \sqrt{-\frac{\Delta\chi^2}{2}}.
\end{align}
\end{ceqn}

The value of this cut was chosen based on injection tests and estimates of the expected false-positive and false-negative rates (see Appendix \ref{sec:TPFP}). It is also clear from Figure \ref{fig:chi2_tot} that this cut removes most of the spurious negative $EW_{\rm{DIB}}$ (emission-like rather than absorption-like) features.\footnote{While DIB emission is not impossible, the only claimed example of emission related to DIB carriers is controversial and in the extreme conditions of the Red Rectangle nebula \citep{Lai_2020_MNRAS}. Thus, DIB emission is not likely present in the sample considered here.} After this choice of DIB SNR cut, 7,789 spectra remain. For any choice DIB SNR cut, catalog impurity and noise is worse at low SNR. We view this DIB SNR cut as a lower bound that should be increased to obtain higher-quality samples as desired. The DIB SNR provides a well-motivated continuous scale on which to make such quality cuts.

The catalog created after these cuts is shown in Figure \ref{fig:cat_dist} as 2D histograms in the space of the moment-0, moment-1, moment-2, and DIB SNR of the detections. Clearly, the vast majority of the catalog has the expected (and likely true) positive absorption and moderate $\sigma_{\rm{DIB}} \sim 1.9$ \r{A}. Further, Figure \ref{fig:cat_dist} illustrates that spurious negative $EW_{\rm{DIB}}$ detections are only a small fraction of the catalog and are only present at low DIB SNR. This explicitly demonstrates how DIB SNR can be use as a well-motivated quality cut.

\subsection{Injection Tests} \label{sec:Inject}

To quantify biases and validate reported uncertainties in both the component separation and catalog creation steps, we created a set of injection tests. To generate these tests we start from a ``parent'' set of sources along lines of sight with low-reddening, SFD $< 0.05$ mag \citep{Schlegel_1998_ApJ}, and with high-quality upstream processing (GSP-Spec flags 1-13 equal to zero). Then, we randomly draw a position, width, and amplitude ($\lambda_{\rm{DIB}}$, $\sigma_{\rm{DIB}}$, $\alpha_{\rm{DIB}}$) as well as an index of a ``parent'' spectrum for the synthetic DIB. Each of these parameters are uniformly sampled over a wide range:  $\lambda_{\rm{DIB}} \in [8618.47, 8628.47]$ \r{A}, $\sigma_{\rm{DIB}} \in [0.6, 3.1]$ \r{A}, $\alpha_{\rm{DIB}} \in [0.25, 0.0]$ \r{A}. 

By uniformly sampling over ``parent'' spectra, the injection tests approximately follow the distribution of stellar SNR in the overall sample (though the SFD and GSP-Spec flag requirements modify the distribution). Since DIB absorption is multiplicative, absorbing the background star light, to obtain the additive DIB component we must multiply the Gaussian profile with the drawn parameters by an estimate of the stellar component for the ``parent'' source. Since we have a stellar component estimate from a star+residuals model (see Section \ref{sec:Model}), we use that stellar component for the ``parent'' source. Then, the synthetic spectrum is the original parent spectrum plus the new DIB component, and we leave the ``parent'' per-pixel uncertainties unchanged.

These injection tests are then subjected to the same pipeline, component separation, and catalog creation cuts, as the real spectra. Given a choice of what a ``good'' detection is, we can compute a measure of the catalog completeness as a function of DIB parameters (see Appendix \ref{sec:TPFP}). Since the injection tests have a ground truth, we can compute Z-scores, the predicted minus true value, divided by the reported uncertainty. When the width of that distribution is one, the reported uncertainties are correctly estimated. Wider than one indicates underestimation and narrower than one indicates overestimation of the reported error bars. 

Figure \ref{fig:inject} shows the Z-score distributions for the three DIB moments ($EW_{\rm{DIB}}$, $\lambda_{\rm{DIB}}$, $\sigma_{\rm{DIB}}$) as a function of the ground truth parameters, stellar SNR, and DIB SNR. As a function of injected $EW_{\rm{DIB}}$, the estimated $EW_{\rm{DIB}}$ Z-score distribution is close to uniform, except at low injected $EW_{\rm{DIB}}$ where a positive bias is observed. The cause of this bias is the imposition of a detection threshold, as evidenced by the stronger bias at low DIB SNR (see Section \ref{sec:LowSNR} for more). The $EW_{\rm{DIB}}$ Z-score distribution is almost perfectly uniform with respect to the injected $\lambda_{\rm{DIB}}$ and $\sigma_{\rm{DIB}}$. There is little bias in the $EW_{\rm{DIB}}$ estimate and error bars as a function of stellar SNR where there are a high density of samples and moderate stellar SNR. At low stellar SNR, there is a $\sim0.5\sigma$ positive bias and underestimation of the error bars. This is likely because spectra with low stellar SNR tend to have lower DIB SNR.

The estimated $\lambda_{\rm{DIB}}$ distribution is almost perfectly uniform with respect to injected $EW_{\rm{DIB}}$, $\lambda_{\rm{DIB}}$, $\sigma_{\rm{DIB}}$, and stellar SNR. As a function of DIB SNR, it is uniform except for at the lowest SNR, where there is a small ($\leq50\%$) underestimation of the error bars.

The estimated $\sigma_{\rm{DIB}}$ distribution is almost perfectly uniform with respect to injected $EW_{\rm{DIB}}$, $\lambda_{\rm{DIB}}$, and $\sigma_{\rm{DIB}}$, but shows a slight ($\leq 0.25\sigma$) positive bias at large injected $EW_{\rm{DIB}}$ and low injected $\sigma_{\rm{DIB}}$. This is the result of a detection bias on $\sigma_{\rm{DIB}}$, which is less commonly known compared to the $EW_{\rm{DIB}}$ (or flux) detection bias (see Section \ref{sec:LowSNR}). At low DIB SNR, and low stellar SNR which tend to have lower DIB SNR, a larger positive bias on the order of $0.5\sigma$ is observed.

\begin{figure}[t]
\centering
\includegraphics[width=\linewidth]{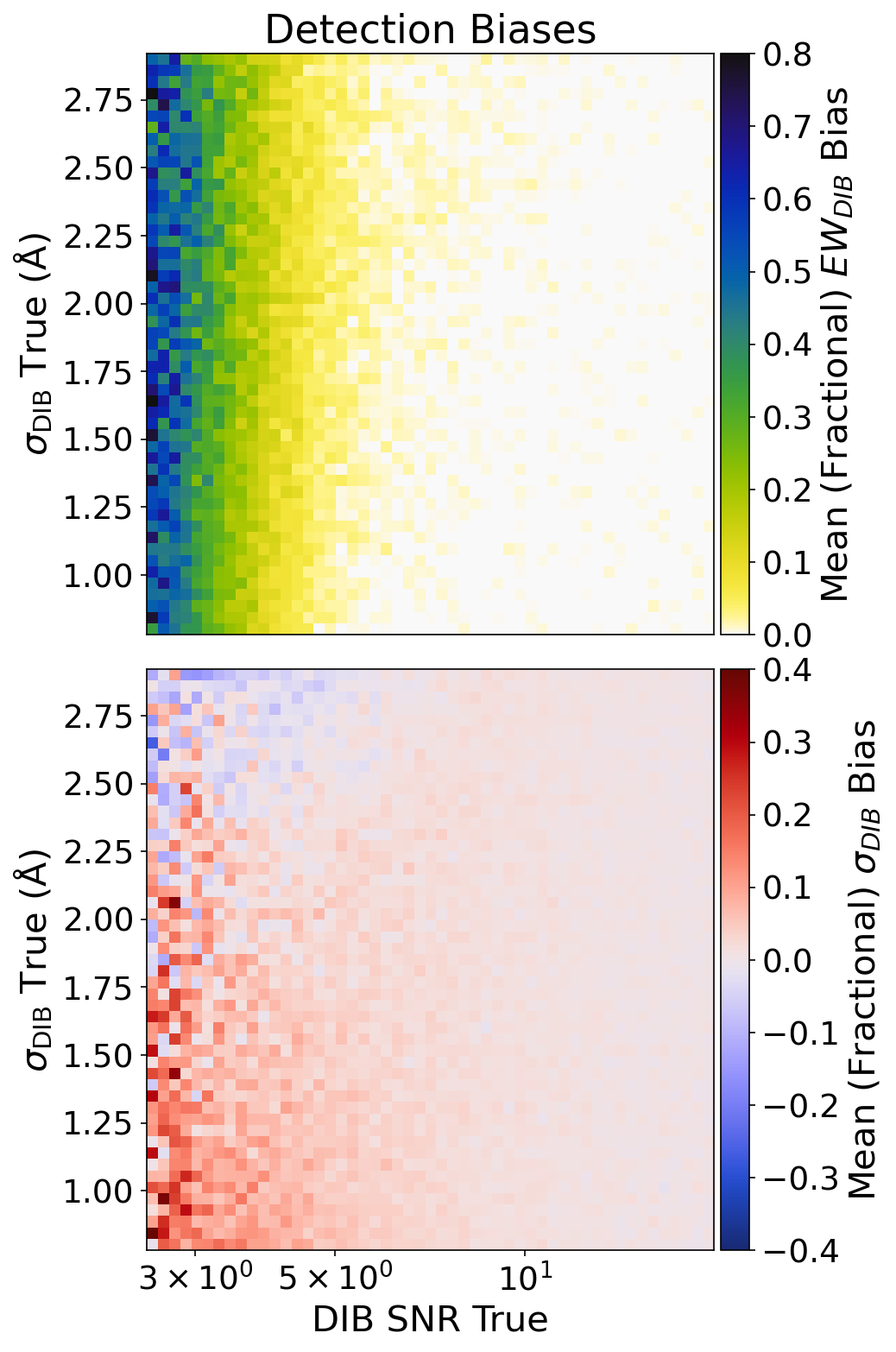}
\caption{Fractional detection bias determined by injection tests for $EW_{\rm{DIB}}$ (top) and $\sigma_{\rm{DIB}}$ (bottom) as a function of the true DIB SNR and true $\sigma_{\rm{DIB}}$.
}
\label{fig:lowSNRemp}
\end{figure}

Figure \ref{fig:inject} is intended primarily as a validation of the method and gives a slightly optimistic view of biases in the final catalog because it uses injections performed on high quality spectra. An injection test releasing the GSP-Spec flag condition is explored in Appendix \ref{sec:TPFP}. On this sample that is more representative of the spectral diversity in the catalog, errors can be $\sim10\%$ underestimated and biases are slightly elevated (see Section \ref{sec:dataavil}). 

Overall, the injection tests allow us to demonstrate impressive uniformity in the recovered Z-score distributions which lends confidence to the parameter estimates and uncertainties in the MADGICS DIB catalog.

\subsection{Low SNR Behavior} \label{sec:LowSNR}

Injection tests subjected to the same detection cuts as the catalog provide an empirical measure of the biases imposed by the creation of a catalog (Figure \ref{fig:lowSNRemp}). Understanding this detection bias function aids in interpreting population-level results from the catalog and can be incorporated in downstream inference. Inference wishing to avoid biases imposed by the detection cuts should work further upstream of the detection cuts (components of all spectra are available in Section \ref{sec:dataavil}). 

Toward low true DIB SNR, we see the expected positively diverging detection bias causing an overestimation of $EW_{\rm{DIB}}$. The $\sigma_{\rm{DIB}}$ has an overall envelope of increasing magnitude toward low true DIB SNR. However, the $\sigma_{\rm{DIB}}$ goes through a sign change for sufficiently large true $\sigma_{\rm{DIB}}$. This sign change is driven primarily by the implicit prior we have imposed by the MADGICS parameter search over a fixed grid of $\sigma_{\rm{DIB}}$. This grid truncates the larger sigma wing of the true posterior for $\sigma_{\rm{DIB}}$, causing a negative $\sigma_{\rm{DIB}}$ bias. From high SNR measurements, we have a physical prior expectation that most true $\sigma_{\rm{DIB}}$ values lie within 1.5 - 2.3 \r{A} for which the $\sigma_{\rm{DIB}}$ bias is predominantly positive.

The form of the low SNR bias can be reproduced theoretically only by combining the effects of both the fixed $\Delta\chi^2$ detection cut and bias correction for using parameter estimates in a Gaussian formalism for a likelihood that has nonlinear dependence on those parameters \citep{Zanolin_2010_PhRvD, Portillo_2017_AJ}. This is because we are approximating as Gaussian the parameter posteriors that are non-Gaussian due to that nonlinearity. The quantitative theoretical model will be described in a future work.

\vspace{-0.5em}
\section{Data/Code Availability} \label{sec:dataavil}
\vspace{-0.25em}
The final catalog, code that implements MADGICS, and several small summary files are provided via \href{https://doi.org/10.5281/zenodo.7388333}{Zenodo} (2 GB). Jupyter notebooks that generate the priors, injection tests, and the catalogs as well as reproduce all figures are also provided. Individual files contained in the Zenodo, including the final catalog alone (3 MB), are available from the \href{https://faun.rc.fas.harvard.edu/saydjari/GaiaDIB/}{project website}. The quality of the final catalog can be further refined by imposing a cut on the DIB SNR field. The full set of intermediate data products, including the full star+residual components, star+DIB+residual components, and injection tests are also available from the project website (500 GB total).
\vspace{-0.5em}
\section{Conclusion} \label{sec:conc}
\vspace{-0.25em}
In this work, we introduced a new technique based on Marginalized Analytic Data-space Gaussian Inference for Component Separation (MADGICS) to detect and measure the properties of DIBs in stellar spectra. Using MADGICS, we pushed detections to lower stellar SNR and marginalized over stellar features, resulting in a new Gaia RVS 8621\,\r{A} DIB catalog free from detectable stellar contamination. In this catalog, we find no evidence for DIB detections within the Local Bubble.

We present and validate a 4D map of ISM kinematics based on the $1^{\rm{st}}$ \aks{moment} of the DIBs. As part of this work, we adjusted the measured rest-frame wavelength to $8623.14 \pm 0.087$ \r{A}, with a precision accounting for modeling uncertainties. Further, we show unprecedented correlation of the DIBs with kinematic substructure in Galactic CO maps. An interpretable and physically realistic trend in the $2^{\rm{nd}}$ moment of the DIBs was also presented. Rigorous validation of the catalog, its reported uncertainties, and the choices made in constructing it are provided via a series of synthetic injection tests. This allows modeling of the low DIB SNR detection biases in both $EW_{\rm{DIB}}$ and $\sigma_{\rm{DIB}}$.

In future work, MADGICS could be extended to more complicated spectral component separation tasks, including identifying multiple stellar components (and the associated radial velocities), removing sky emission lines (from ground-based observations), and modeling DIB detections with multiple cloud components (velocities) along the line of sight. Looking forward to increased source densities in future data releases, larger high-quality 4D kinematic maps of the ISM will enable detailed studies of ISM substructure and dynamics \citep{Tchernyshyov_2018_AJ}.

\acknowledgments{
A.K.S. gratefully acknowledges support by a National Science Foundation Graduate Research Fellowship (DGE-1745303). D.P.F. acknowledges support by NSF grant AST-1614941, “Exploring the Galaxy: 3-Dimensional Structure and Stellar Streams.” D.P.F. acknowledges support by NASA ADAP grant 80NSSC21K0634 “Knitting Together the Milky Way: An Integrated Model of the Galaxy’s Stars, Gas, and Dust.” C.Z. acknowledges that support for this work was provided by NASA through the NASA Hubble Fellowship grant HST-HF2-51498.001 awarded by the Space Telescope Science Institute (STScI), which is operated by the Association of Universities for Research in Astronomy, Inc., for NASA, under contract NAS5-26555. The support and resources from the Center for High Performance Computing at the University of Utah are gratefully acknowledged. A portion of this work was also enabled by the FASRC Cannon cluster supported by the FAS Division of Science Research Computing Group at Harvard University. 
}

This work was supported by the National Science Foundation under Cooperative Agreement PHY-2019786 (The NSF AI Institute for Artificial Intelligence and Fundamental Interactions). The authors acknowledge Interstellar Institute's program ``With Two Eyes'' and the Paris-Saclay University's Institut Pascal for hosting discussions that nourished the development of the ideas behind this work.

We acknowledge helpful discussions with Mathias Schultheis, He Zhao, Morgan Fouesneau, Jos de Bruijne, Patrick de Laverny, Gordian Edenhofer, Torsten Enßlin, Gail Zasowski, Joel Brownstein, Tom Dame, Greg Green, Eddie Schlafly, Jonathan Bird, Adam Wheeler, Andy Casey, and Phillip Cargile. A.K.S. acknowledges Sophia S\'{a}nchez-Maes for helpful discussions and much support.

\begin{figure*}[t]
\centering
\includegraphics[width=\linewidth]{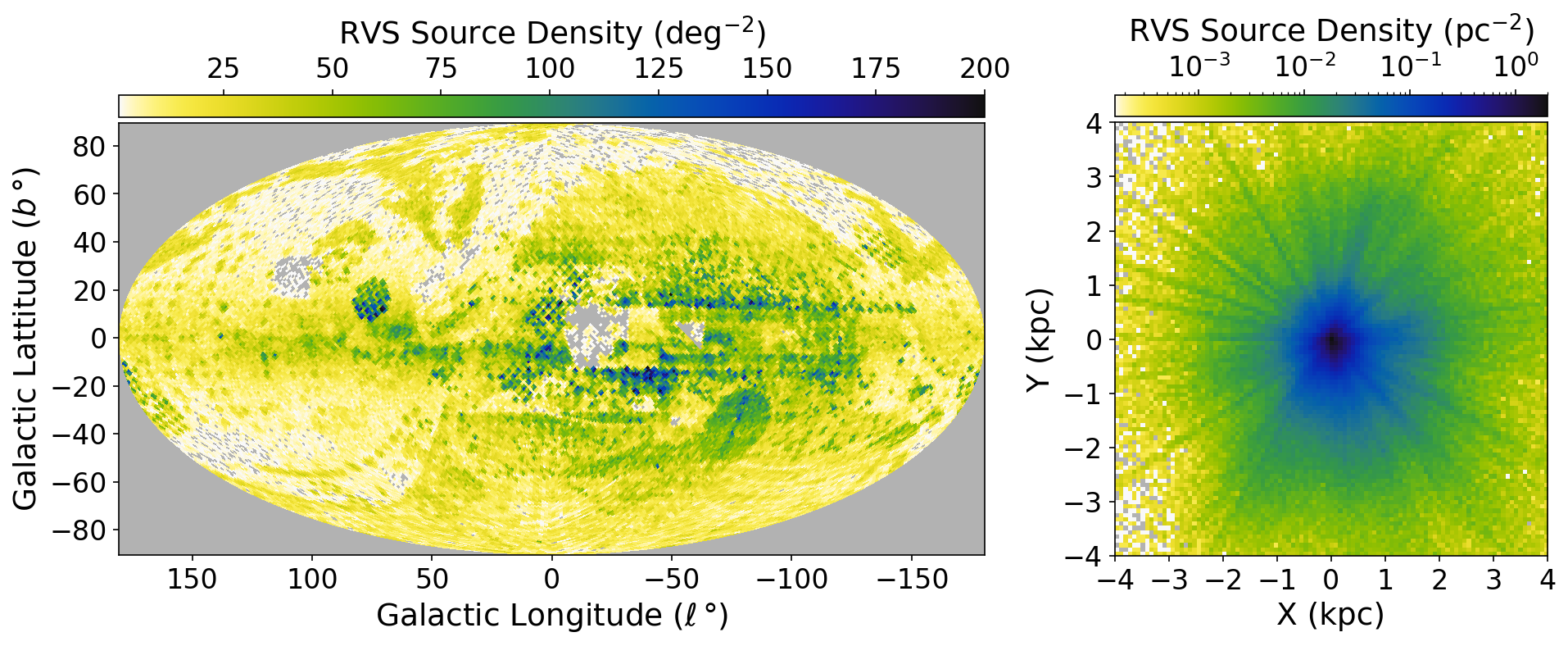}
\caption{Density of public RVS spectra on the sky (left, HEALPix grid at NSide = 64) and projected onto the Galactic plane in Cartesian coordinates (right).}
\label{fig:selectFun}
\end{figure*}

This work has made use of data from the European Space Agency (ESA) mission
{\it Gaia} (\url{https://www.cosmos.esa.int/gaia}), processed by the {\it Gaia}
Data Processing and Analysis Consortium (DPAC,
\url{https://www.cosmos.esa.int/web/gaia/dpac/consortium}). Funding for the DPAC
has been provided by national institutions, in particular the institutions
participating in the {\it Gaia} Multilateral Agreement. This research has made use of the VizieR catalog access tool, CDS, Strasbourg, France \citep{Ochsenbein_2000}.

\facility{Gaia}

\software{
Julia \citep{bezanson2017julia},
FITSIO.jl \citep{Pence_2010_A_A},
HDF5.jl \citep{hdf5},
Healpix.jl \citep{2021ascl.soft09028T},
BenchmarkTools.jl \citep{BenchmarkTools.jl-2016},
\textsc{astropy} \citep{AstropyCollaboration:2013:A_A:},
\textsc{ipython} \citep{Perez:2007:CSE:}, 
\textsc{matplotlib} \citep{Hunter:2007:CSE:},
\textsc{scipy} \citep{2020SciPy-NMeth}.
}

\onecolumngrid
\newpage 

\appendix
\begin{appendices}


\begin{figure}[t]
\centering
\includegraphics[width=\linewidth]{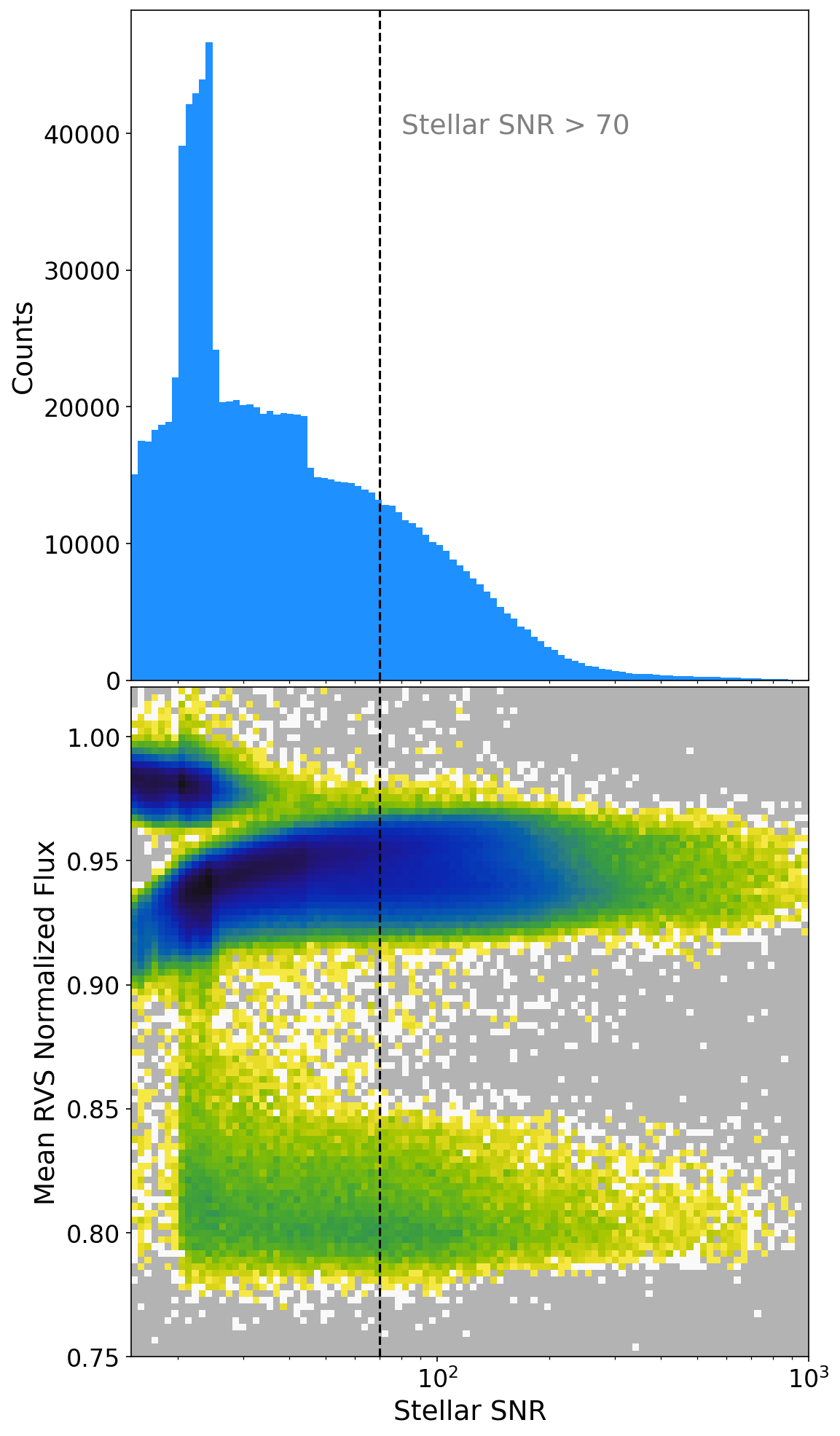}
\caption{(Top) Histogram of public Gaia RVS sources with respect to stellar SNR. (Bottom) 2D histogram of public Gaia RVS sources with respect to stellar SNR and the mean normalized flux in the RVS spectrum. The color scale indicates density, where darker colors indicate more sources, and is a logarithmic stretch. A guideline is shown at stellar SNR 70, which is the cutoff imposed by the Gaia DIB catalog.}
\label{fig:RVSSNR}
\end{figure}

\section{Public RVS Selection Function} \label{sec:selectFun}

We illustrate the distribution of public RVS spectra in Figure \ref{fig:selectFun}. The highly heterogeneous distribution on the sky (shown in Galactic coordinates) shows clear patterns associated with the footprints of other surveys, such as Kepler \citep{Borucki_2010_Sci}, Galactic Archaeology with Hermes (GALAH, \citealt{Buder_2021_MNRAS}), the Large sky Area Multi-Object Fiber Spectroscopic Telescope (LAMOST, \citealt{Luo_2015_RAA}), and the Apache Point Observatory Galactic Evolution Experiment (APOGEE, \citealt{Majewski_2017_AJ}). We understand this to be because the majority of the public RVS spectra were those used for validation by comparing to other spectroscopic measurements of the same source \aks{\citep{Vallenari_2022_arXiv, Seakbrokeunpub}.}

Heterogeneity is also seen with respect to the stellar SNR, as shown in Figure \ref{fig:RVSSNR}. Plotting as a function of stellar SNR also highlights the significant heterogeneity in the normalization of the RVS spectra as a result of a discrete choice in the Gaia pipeline to either pseudo-continuum normalize or rescale the spectra by a constant (in the case of cool stars or low stellar SNR) \citep{Vallenari_2022_arXiv}. 

\section{DIB SNR Distribution} \label{sec:DIBSNR}

\begin{figure}[t]
\centering
\includegraphics[width=0.94\linewidth]{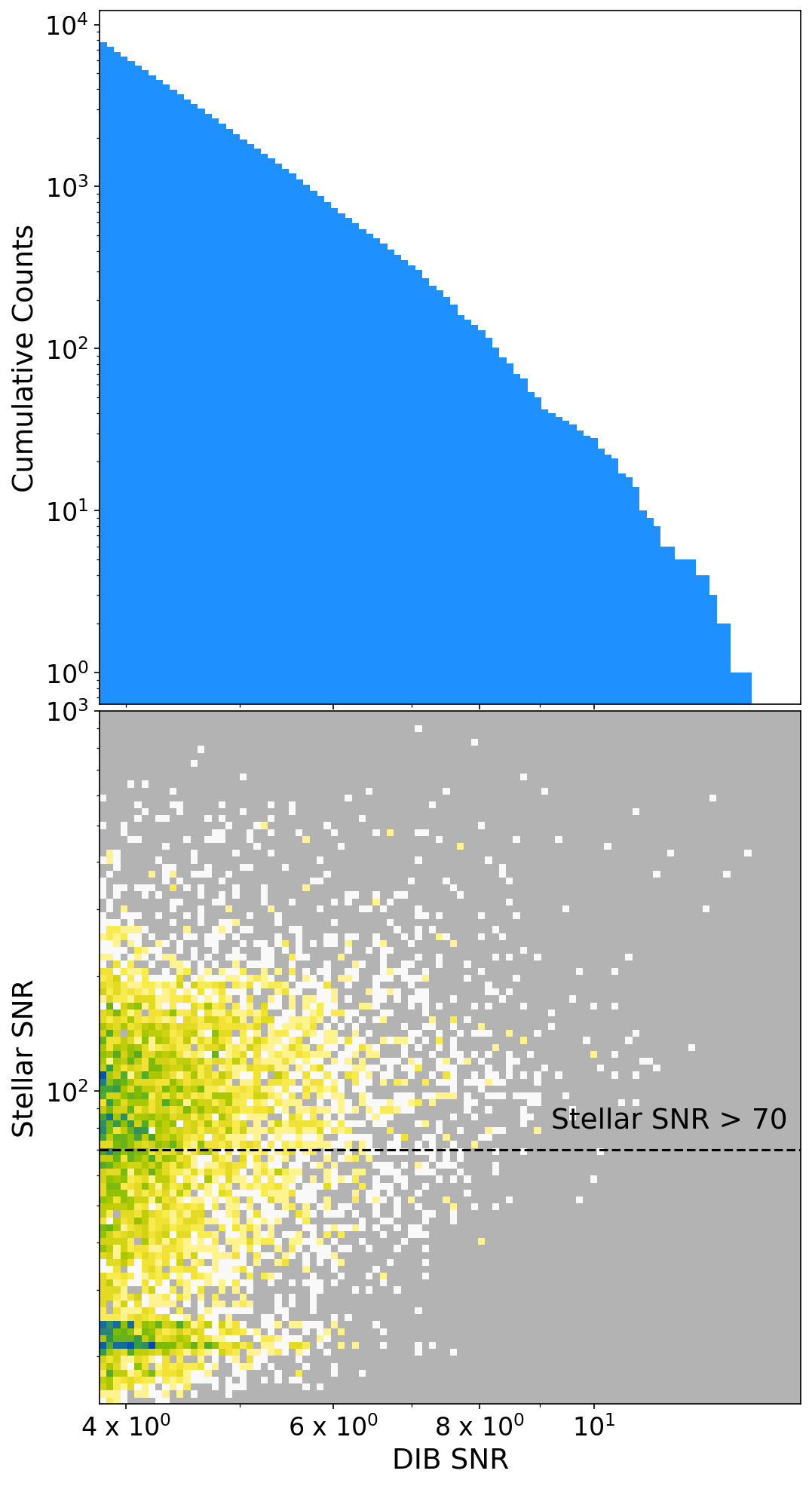}
\caption{(Top) Cumulative counts of MADGICS DIB detections as a function of DIB SNR. (Bottom) 2D histogram of MADGICS DIB detections as a function of DIB and stellar SNR. The color scale indicates density, where darker colors indicate more sources, and is a logarithmic stretch. A guideline is shown at stellar SNR 70, which is the cutoff imposed by the Gaia DIB catalog.}
\label{fig:CumulSNR}
\end{figure}

It is useful to describe the MADGICS DIB catalog detections as a function of SNR. In the top of Figure \ref{fig:CumulSNR}, we show the cumulative detections as a function of the reported DIB SNR. This provides an idea of the number of detections remaining in the catalog for some choice of quality cut using the DIB SNR. In the bottom of Figure \ref{fig:CumulSNR}, we show the 2D histogram of the detections as a function of both DIB SNR and stellar SNR. Clearly, many high-quality detections (large DIB SNR) were found below the stellar SNR cutoff imposed by the Gaia DIB catalog. Because a large fraction of the spectra have low SNR, it is important to leverage their statistical power. The peak at stellar SNR $\sim20$ is simply a reflection of the unusually large number of spectra at that stellar SNR in the public Gaia DR3 RVS spectra. We have confirmed that the fraction of DIB detections passing our DIB SNR cut actually decreases as a function of stellar SNR. This lends confidence that MADGICS is correctly finding the large DIB features which have high DIB SNR even on low stellar SNR spectra, without overfitting noise as DIB detections.

\section{\aks{Removing} Goodness of Fit Cuts} \label{sec:noGOF}

\begin{figure*}[t]
\centering
\includegraphics[width=\linewidth]{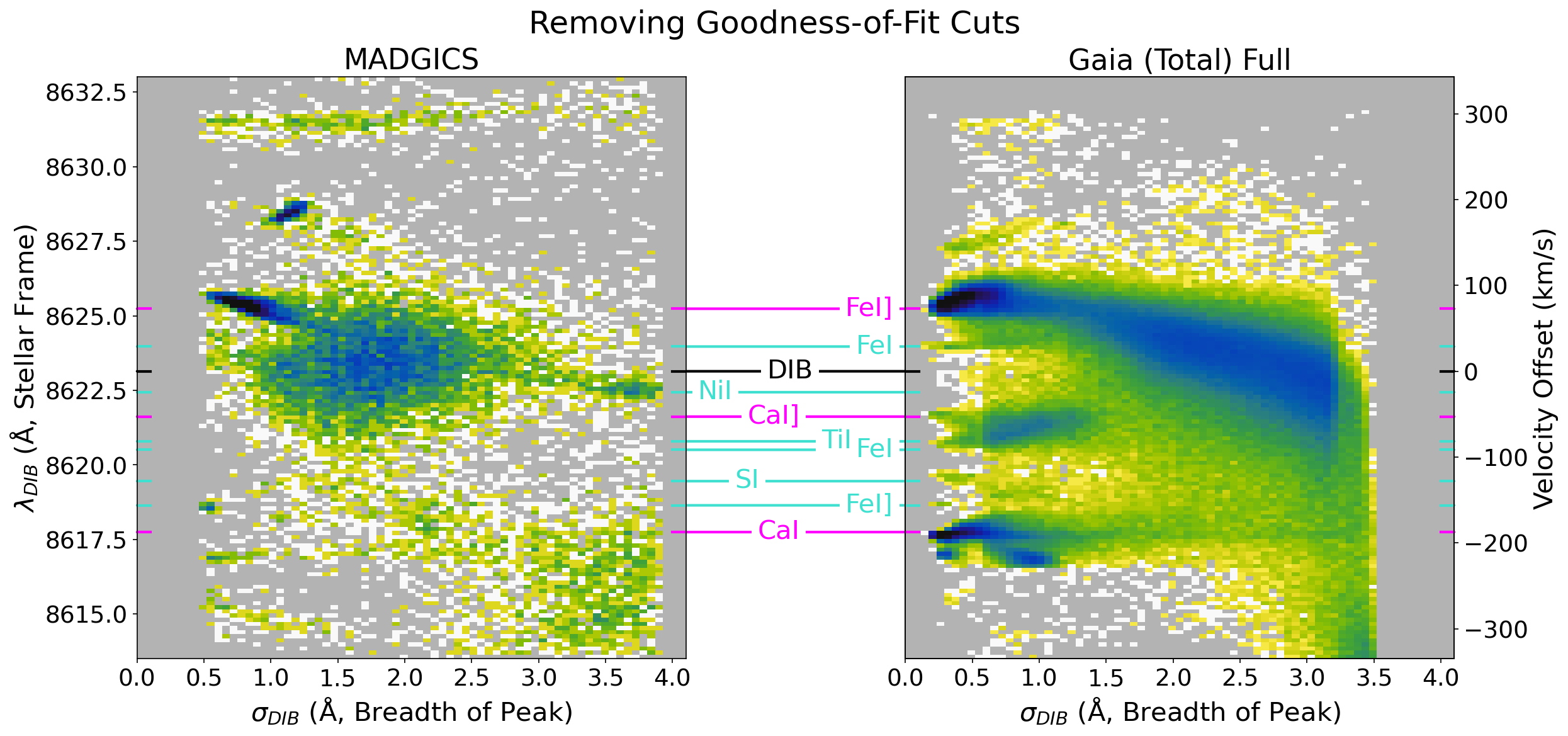}
\caption{Density of detections in the MADGICS (left) and Gaia (right) 8623\r{A} DIB catalog as a function of $\lambda_{\rm{DIB}}$ and $\sigma_{\rm{DIB}}$, \aks{removing} cuts imposed on the goodness of fit of the model to the overall spectrum. \aks{This removes the second block of cuts in Table \ref{tab:dibcat} for both catalogs.} The Gaia catalog shown is the full DIB sample, including \aks{DIB detections in the public Gaia DR3 DIB catalog that were} identified on spectra that are not public. The color scale indicates density, where darker colors indicate more detections, and is a relative logarithmic stretch for both catalogs, with the maximum rescaled by relative catalog sizes. Guidelines for the central wavelength of the DIB and stellar lines are labeled and color coded between the plots.}
\label{fig:contamfignoGOF}
\end{figure*}

\begin{figure}[b]
\centering
\includegraphics[width=0.96\linewidth]{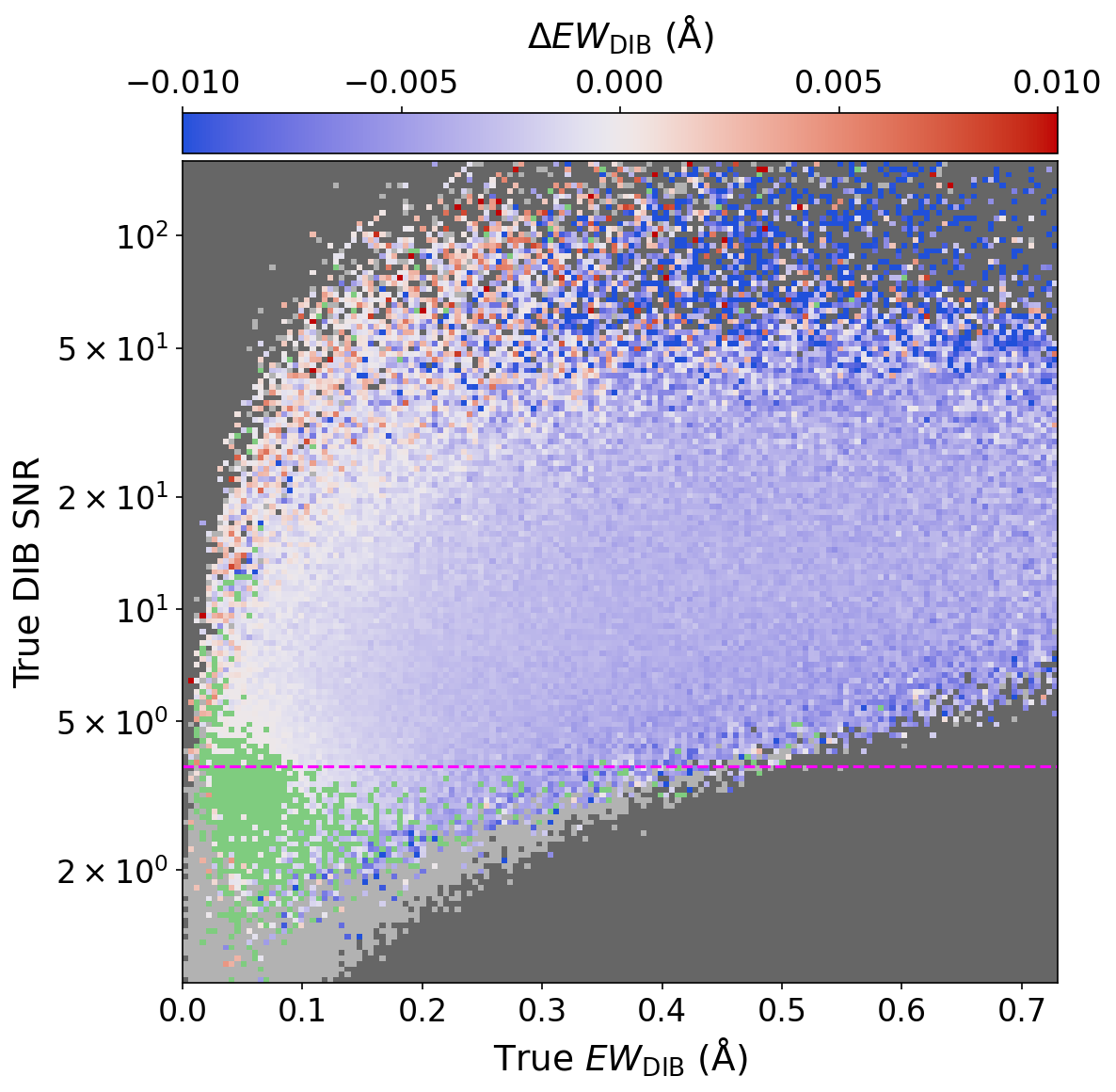}
\caption{Median change in $EW_{DIB}$ as a function of $EW_{DIB}$ and DIB SNR after including dustier lines of sight in the training data for the stellar covariance. Green indicates $>25\%$ of detections not found, gray indicates detections not found with the clean stellar covariance, and dark gray indicates no injection tests present. The guideline (magneta, dashed) indicates the detection cut on estimated DIB SNR impose in making the MADGICS catalog.}
\label{fig:dustyCat}
\end{figure}

As stated in the main text, we aim to (1) impose goodness of fit cuts that eliminate any influence of stellar lines and (2) use methods that are maximally robust to stellar mismodeling so that real DIBs can be detected and well-modeled in the largest number of spectra possible. To help compare the catalogs with respect to (2), we replicate Figure \ref{fig:contamfig} without any goodness of fit cuts imposed. Without goodness of fit cuts, MADGICS would return detections in 19,071 spectra compared to the 50,787 detections in the Gaia Full catalog on public spectra.

Clearly, there are detections associated with several of the stellar lines identified in Figure \ref{fig:contamfig} within the MADGICS detections when goodness of fit cuts are not imposed. This is to be expected. When the star is severely mismodeled, removing the stellar component and finding the broad multiplicative absorption of the DIB is nearly impossible. The question is then ``How robust is the DIB identification to the stellar mismodeling?'' The lower density of detections associated with stellar features in the MADGICS catalog panel of Figure \ref{fig:contamfignoGOF} suggests that the MADGICS approach is also more robust (2), in addition to having the high-quality goodness of fit cut (1) demonstrated in Section \ref{sec:StellarFrame}.

\section{Dusty Stellar Prior} \label{sec:DustyPrior}

It is difficult to prove that a data-driven stellar model (the stellar covariance herein) does not have any small contributions from low $EW_{DIB}$ lines of sight that are present in the training data despite the cut on SFD $< 0.05$. However, we can easily show the impact of the inverse, where we loosen the cut to SFD $< 0.1$ during the construction of the stellar prior, thereby including much dustier lines of sight that are likely to have more contributions from low $EW_{DIB}$ DIBs. We then use this ``dusty'' stellar prior to detect DIBs on the synthetic injection tests described in the main text, and show the resulting change in the DIB catalog in Figure \ref{fig:dustyCat}.

Figure \ref{fig:dustyCat} shows the median change in $EW_{\rm{DIB}}$ as a function of the true $EW_{\rm{DIB}}$ and true DIB SNR of the injected DIBs. We take this difference between DIB detections appearing in both the original and ``dusty stellar prior'' DIB catalogs built from the injection tests. Regions of parameter space where $>25\%$ of detections in the original catalog were not refound are colored green. Only a very small number (557/1,000,000) of injections were detected in the ``dusty stellar prior'' catalog, but not the original and are excluded from this plot. Regions of parameter space where no DIBs were detected in either catalog are shown in light gray, and regions not probed by the injection tests are shown in dark gray.

Over the majority of the detections, there is a $<0.005$ \r{A} reduction in the reported $EW_{\rm{DIB}}$. Just near the detection threshold, a $<0.005$ \r{A} increase is observed. This can be interpreted as a shift in the detection threshold carrying with it the usual positive detection bias (see Section \ref{sec:LowSNR}). For moderate $EW_{\rm{DIB}}$ ($>0.15$ \r{A}), there is very little change in the detectability of the synthetic injections, even at low DIB SNR. At low $EW_{\rm{DIB}}$ ($<0.15$ \r{A}), there is a clear increase in the true DIB SNR required in order for an injected DIB to be detected. This makes sense as some small contribution of the DIB peak is present in the stellar model, though averaged over the relative velocity distribution between the star and dust. However, almost all of the DIBs that are lost have a true DIB SNR below the DIB SNR cut imposed in making the catalog, and thus were only detected previously because they fluctuated high. Further, it is important to note the persistence of low $EW_{\rm{DIB}}$ detections at higher DIB SNR. 

Resolving the question of the presence of DIBs in the Local Bubble is important. In this work, we show that approaches with synthetic models can incorrectly identify stellar features as DIBs. However, it is difficult to prove that a data-driven stellar model is completely free of contributions from low $EW_{\rm{DIB}}$ DIBs. Yet, Figure \ref{fig:dustyCat} makes it clear that data-driven stellar models can solve this question if we push to higher stellar SNR spectra, since we can then push confident DIB detections to lower $EW_{\rm{DIB}}$ limits.

\section{Dust-CO Correlation} \label{sec:dust_corr}

\begin{figure}[b]
\centering
\includegraphics[width=\linewidth]{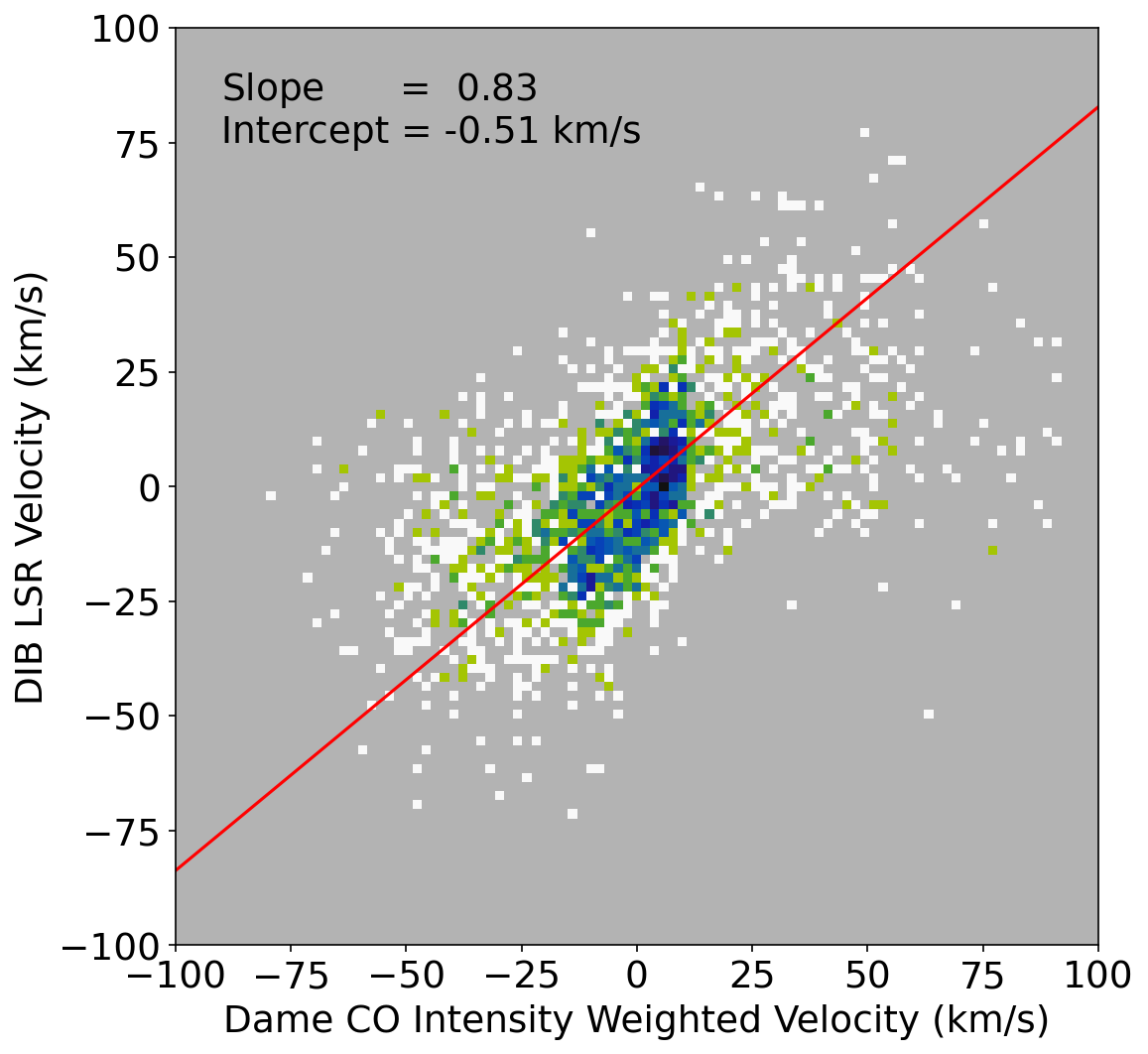}
\caption{Histogram (2D) illustrating correlation between the $v_{\rm{DIB}}^{\rm{LSR}}$ and intensity-weighted velocity of CO over all velocity channels in the \citealt{Dame_2001_ApJ} CO map. The color scale indicates density, where darker colors indicate more detections and a logarithmic stretch has been applied.}
\label{fig:lowVCO}
\end{figure}

To account for the clumpiness of CO, we find the nearest (in velocity) peak in CO to the DIB peak before calculating the intensity-weighted mean CO velocity $v_{CO}$ that corresponds to that DIB. First we restrict to only DIBs within the footprint of the \citealt{Dame_2001_ApJ} CO survey (3602). Further, we restrict to DIBs such that there is at least some detection of CO within $1\sigma$ of the DIB velocity (2272). Then, we find the nearest connected set of velocity channels with CO detections above threshold, where the threshold starts at zero, and the threshold increases in steps of 0.05 until the connected region (island) is less than 9 velocity pixels wide. After this stopping condition is reached, we take the intensity weighted average over the island to obtain $v_{CO}$.

While such a peak-finding algorithm will necessarily overestimate the correlation between the CO and DIBs, we find a similar correlation when restricting to low velocity CO without any peak-finding, only making the two broad cuts outlined above (Figure \ref{fig:lowVCO}). Computing $v_{CO}$ as the intensity-weighted velocity over all velocity channels, we restrict to $|v_{CO}| < 15$ km/s to reduce the multimodality of the CO and DIB distributions. Then, we find a slope of $0.83 \pm 0.04$ and an intercept of $-0.51 \pm 0.3$ km/s. Both of these values are consistent with those discussed in the text, though the slope is slightly flatter and there is larger scatter due to the additional averaging in $v_{CO}$. However, the scatter of residuals/$\sigma(v_{\rm{DIB}}^{\rm{LSR}})$, or Z-scores, was 0.96, indicating that the reported uncertainties well-describe the scatter in the residuals. Thus, we conclude the peak-finding to determine $v_{CO}$ does not significantly change the conclusions and it helps illustrate that strong correlations are present for high velocity DIBs and CO after accounting for the multimodality.

\section{Purity and Completeness} \label{sec:TPFP}

\begin{figure}[b]
\centering
\includegraphics[width=\linewidth]{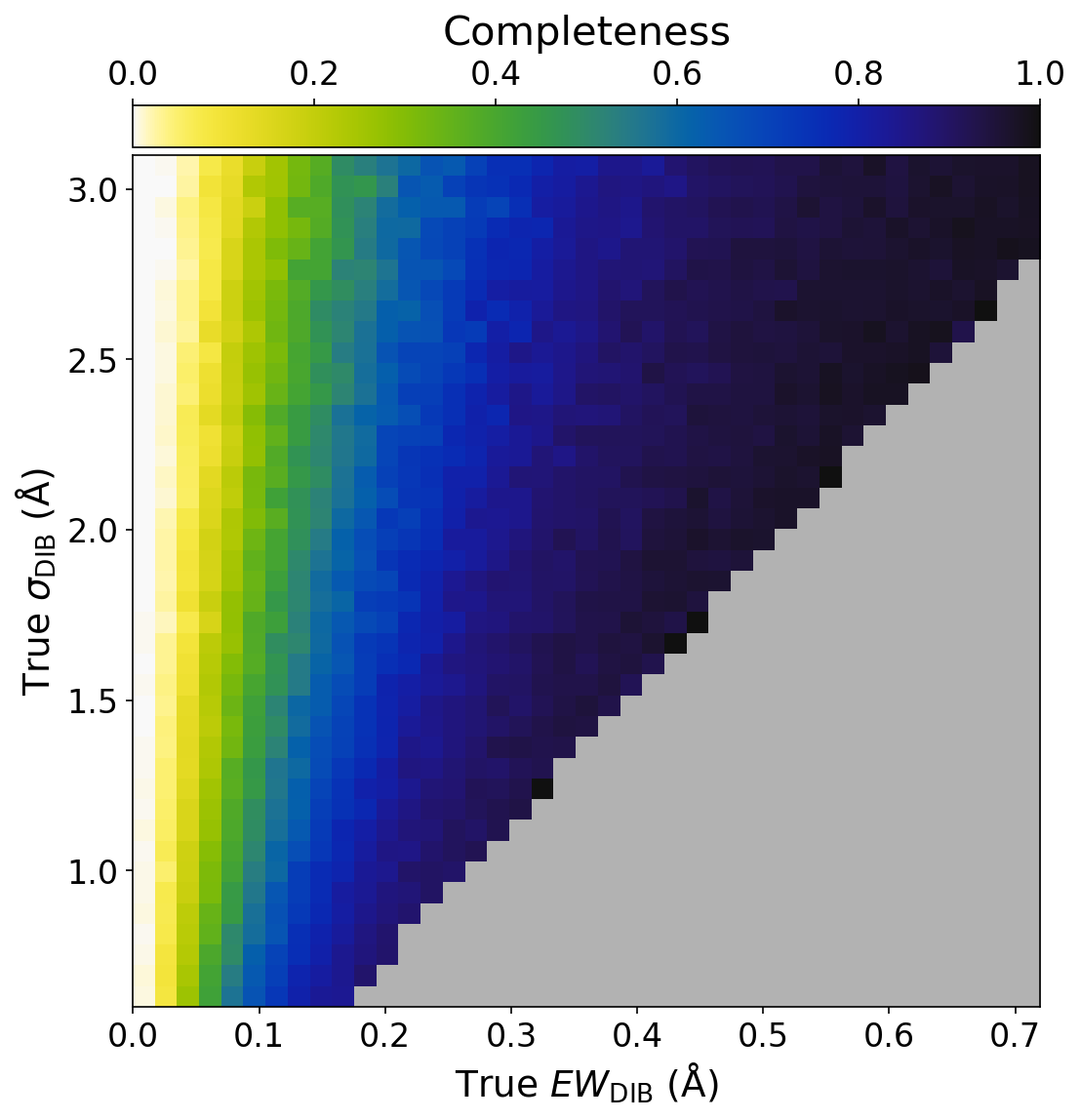}
\caption{Estimated completeness of DIB catalog as a function of $EW_{\rm{DIB}}$ and $\sigma_{\rm{DIB}}$ based on quality cuts on injection tests.}
\label{fig:complete}
\end{figure}

Combining injection tests with a choice for what a ``good'' detection is allows an estimate of the completeness of the catalog (Figure \ref{fig:complete}). Here, we take an arbitrary choice for good detections to be  $|\Delta\alpha_{\rm{DIB}}|/\alpha_{\rm{DIB}} < 0.2$, $|\Delta\lambda_{\rm{DIB}}| < $ 2.5 \r{A}, $|\Delta\sigma_{\rm{DIB}}|/\sigma_{\rm{DIB}} < 0.2$, where the $\Delta$ indicates differences with respect to the ground truth values for the injected DIB profile. We show this completeness as a function of both $EW_{\rm{DIB}}$ and $\sigma_{\rm{DIB}}$ to illustrate that it is more difficult to detect wider DIBs given the same $EW_{\rm{DIB}}$. For $EW_{\rm{DIB}} \geq 0.2$ \r{A}, the catalog is $\geq80\%$ complete over the range of relevant $\sigma_{\rm{DIB}}$ here. Even though completeness is reduced at lower $EW_{\rm{DIB}}$, we still achieve $\sim20\%$ completeness at 0.05-0.1 \r{A}.

\begin{figure}[t]
\centering
\includegraphics[width=\linewidth]{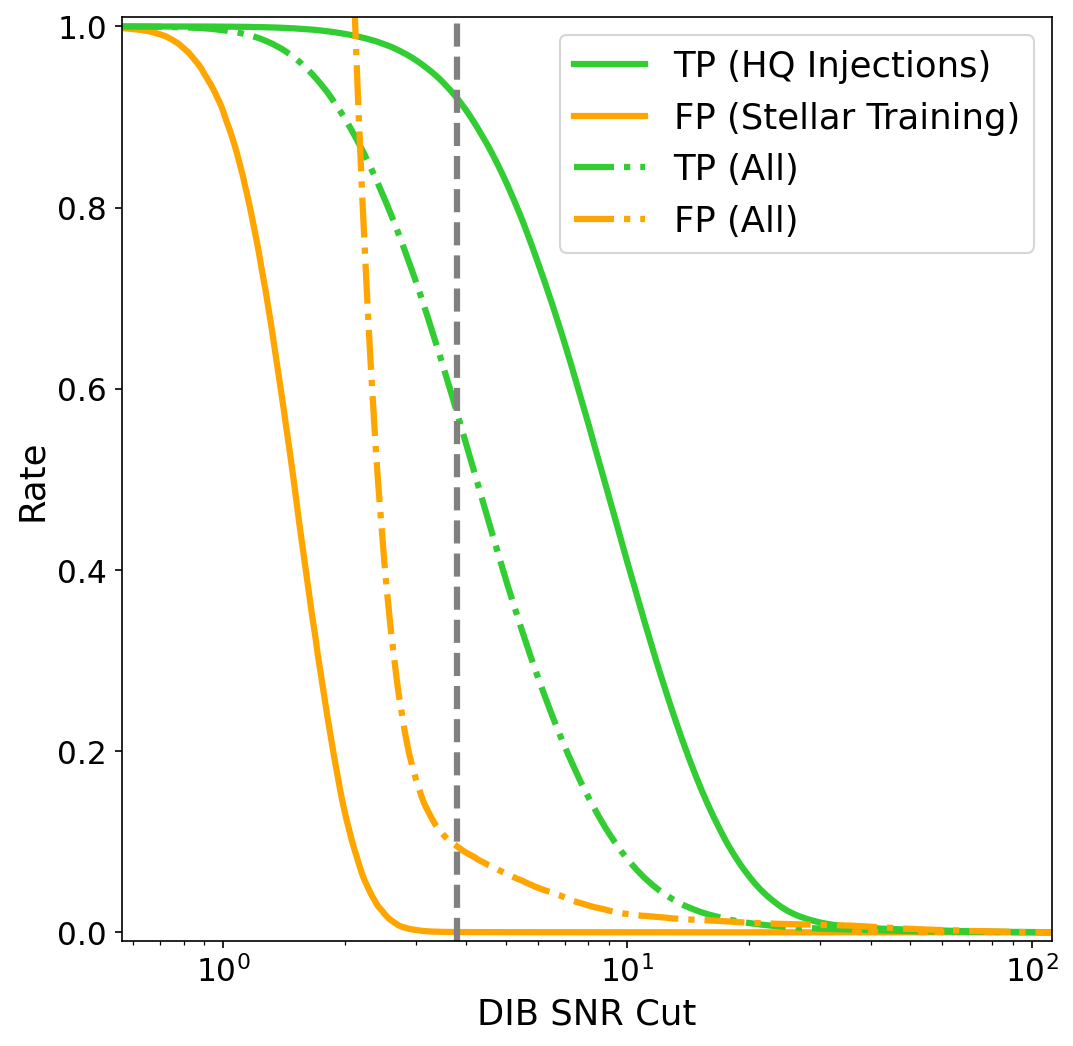}
\caption{Estimated true positive (TP) and false positive (FP) rates as a function of DIB SNR cut using the injection tests and stellar training set, respectively (solid lines). Estimated TP and FP rates using a broader set of low reddening injection tests/spectra are shown in dot-dashed lines. Vertical guideline illustrates value chosen in this work (gray, dashed).}
\label{fig:TPFP}
\end{figure}

In creating Figure \ref{fig:complete}, we are treating the detection of synthetic injections as ``true positives'' (TP). We can leverage this to help motivate a choice for the $\Delta\chi^2$ cut used to make the catalog. To do this, we compute the TP rate as a function of the $\Delta\chi^2$ cut chosen, restricting to $\alpha_{\rm{DIB}} > 0.07$ as motivated by Figure \ref{fig:complete}. Similarly, we can use the training set used for generating the stellar prior (low reddening, high stellar SNR, no GSP-Spec flags) as a set of ground truth negatives. Then, ``detections'' of a DIB on those spectra give an idea of the ``false positive'' (FP) rate. We can then produce the FP rate as a function of the $\Delta\chi^2$ cut chosen, again restricting to $\alpha_{\rm{DIB}} > 0.07$.

Both of these reference datasets are somewhat idealized. To better capture how the heterogeneity in the full dataset impacts the TP rate, we created a synthetic injection test using a ``parent'' set that only requires SFD $< 0.5$ mag (low-dust) and that the spectra not have been used in the stellar prior training set. Similarly, we can estimate a more realistic FP rate using all low reddening spectra (without an injection step).

In Figure \ref{fig:TPFP} we show these TP and FP rates as a function of the $\Delta\chi^2$ cut chosen, which we have converted into a cut on the DIB SNR. The value used in this work is shown by a dashed vertical line. This corresponds to (FP, TP) = ($0.04\%$,$92\%$) for the high-quality spectra tests and (FP, TP) = ($9\%$,$57\%$) for tests more representative of the average spectra in the Gaia DR3 RVS release. The exact choice of the test sets and definitions of TP and FP will modify the exact rates, but Figure \ref{fig:TPFP} demonstrates the qualitative choice we have made to optimize for low false positive rates. 

\begin{deluxetable}{cccc}[h]
\tablenum{2}
\tablecaption{Stellar Line Labels
\label{tab:linlab}}
\tablecolumns{4}
\tablehead{
Species & Wavelength (\r{A}) & $\log(gf)$ & Included?
}
\startdata
\multicolumn{4}{c}{\emph{Calibrated GSP-Spec Lines}}\\
\hline
\FeI & 8614.169 & -1.850 & Y \\
\FeI & 8616.306 & -1.246 & Y \\ 
\FeI] & 8618.647 & -0.707 & Y \\ 
\SI & 8619.457& -0.763 & Y \\ 
\FeI] & 8620.508 & -1.868 & Y \\ 
\TiI & 8620.792 & -1.980 & Y \\ 
\NiI] & 8622.439 & -1.376 & Y \\
\FeI & 8623.969 & -2.320 & Y \\ 
\SI & 8628.911 & -0.894 & Y \\ 
\SiI] & 8631.971 & -3.168 & Y \\ 
\hline
\multicolumn{4}{c}{\emph{Lines not Calibrated or Included in GSP-Spec}}\\
\hline
$^{\bullet}$CN & 8615.040 & -1.345 & ? \\ 
\FeI & 8617.678 & -0.950 & N \\ 
\FeI] & 8621.491 & -1.330 & N \\
$^{\bullet}$CN & 8621.732 & -1.269 & ? \\ 
\FeI & 8625.112 & -0.770 & N \\ 
\FeI] & 8626.109 & -2.089 & Y \\
\FeI] & 8627.260 & -3.177 & Y \\
$^{\bullet}$CN & 8628.620 & -1.208 & ? \\ 
$^{\bullet}$CN & 8631.382 & -1.104 & ? \\ 
\enddata
\end{deluxetable}\vspace{-12mm}

\section{Line Labels} \label{sec:linelist}

As discussed in Section \ref{sec:stellarprior}, we labeled the pile-ups in the stellar rest-frame wavelength found in the Gaia DIB catalog by the transition at those locations with the largest amplitude in models of the solar spectrum using the Kurucz archive (\href{http://kurucz.harvard.edu/}{http://kurucz.harvard.edu/}). In Table \ref{tab:linlab} we list these lines along with the set of stellar lines in the GSP-Spec model carefully calibrated by \citealt{Contursi_2021_A_A} in the wavelength range near the DIB feature. Of the lines that were not calibrated as part of the GSP-Spec modeling, we were able to confirm that two of the \FeI] lines were part of the (private) GSP-Spec atomic line list \citetext{private communication, P. de Laverny, 2023}. However, the molecular line list is proprietary and thus the inclusion of the $^{\bullet}$CN lines in the GSP-Spec line list remains unknown to us \citetext{private communication, P. de Laverny, 2023}. However, we reiterate that regardless of the inclusion of any given line in the line list, the Gaia DIB catalog clearly has contamination from poorly-modeled stellar features centered on these wavelengths.

\section{Example Full Posteriors} \label{sec:pixCov}

\begin{figure}[h]
\centering
\includegraphics[width=\linewidth]{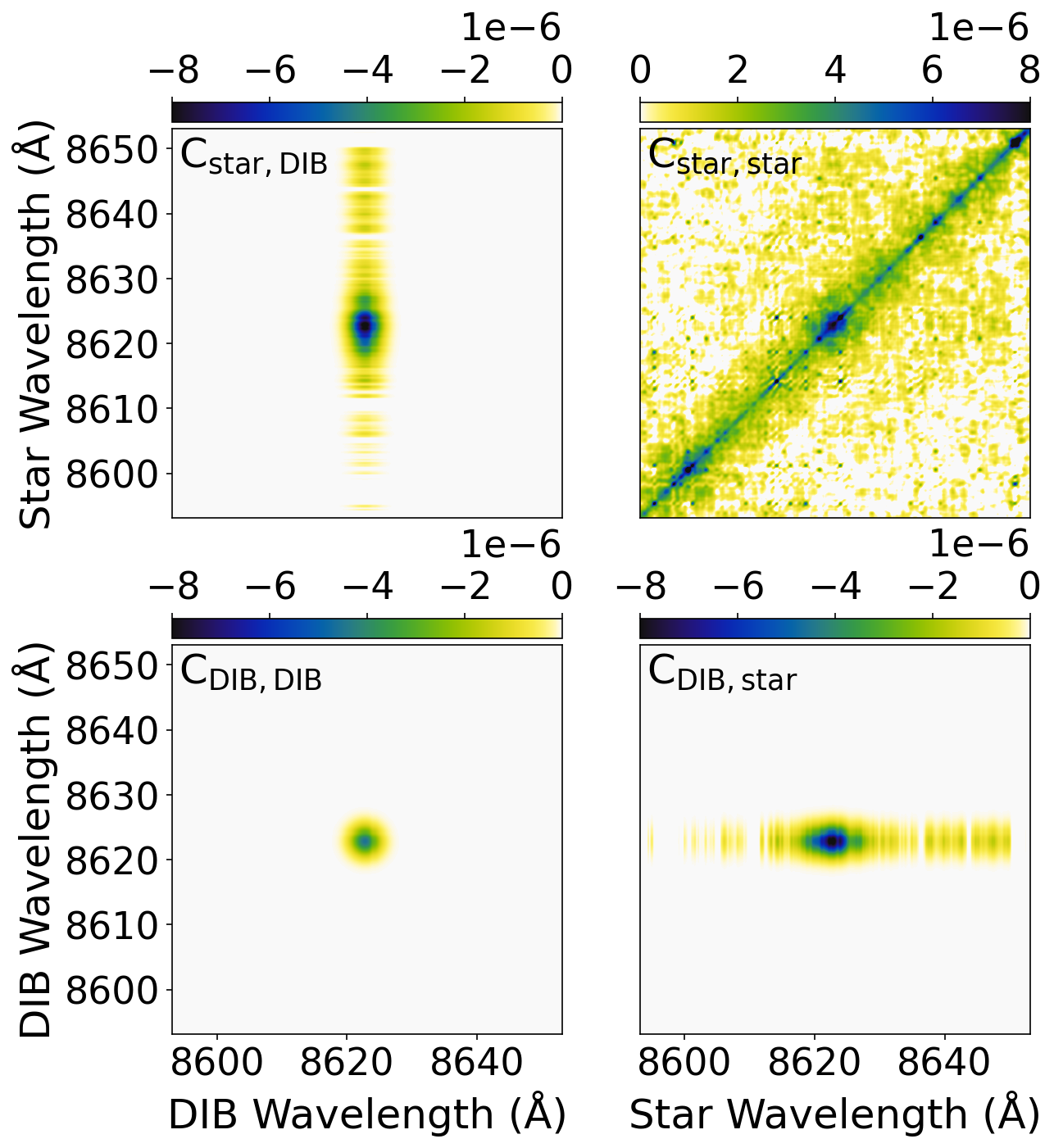}
\caption{\aks{Posteriors for pixels within either the DIB component C$_{\rm{DIB}}$ or stellar component C$_{\rm{stellar}}$. Covariance between pixels spanning the two components are shown as the cross-terms C$_{\rm{DIB},\rm{stellar}}$ and C$_{\rm{stellar},\rm{DIB}}$.}}
\label{fig:cov_sd}
\end{figure}

\aks{As discussed in Section \ref{sec:Stats}, we obtain $\hat{C}_{kk}$, the predicted pixel-pixel covariance of pixels in component $k$, and $\hat{C}_{km}$, the predicted covariance of a pixel in component $k$ with a pixel in component $m$. In Figure \ref{fig:cov_sd}, we show a subset of these covariance matrices that illustrate  the covariance for pixels within and between the stellar and dust components for the decomposition of the spectrum shown in Figure \ref{fig:decomp}.

Because MADGICS imposes that the sum of the components be exactly equal to the data, the covariance on the sum of the components must be identically zero. This can be confirmed directly from Equations \ref{eq:post_kk}--\ref{eq:post_km}}.

\section{Velocity Shift} \label{sec:velShift}

\begin{figure}[t]
\centering
\includegraphics[width=\linewidth]{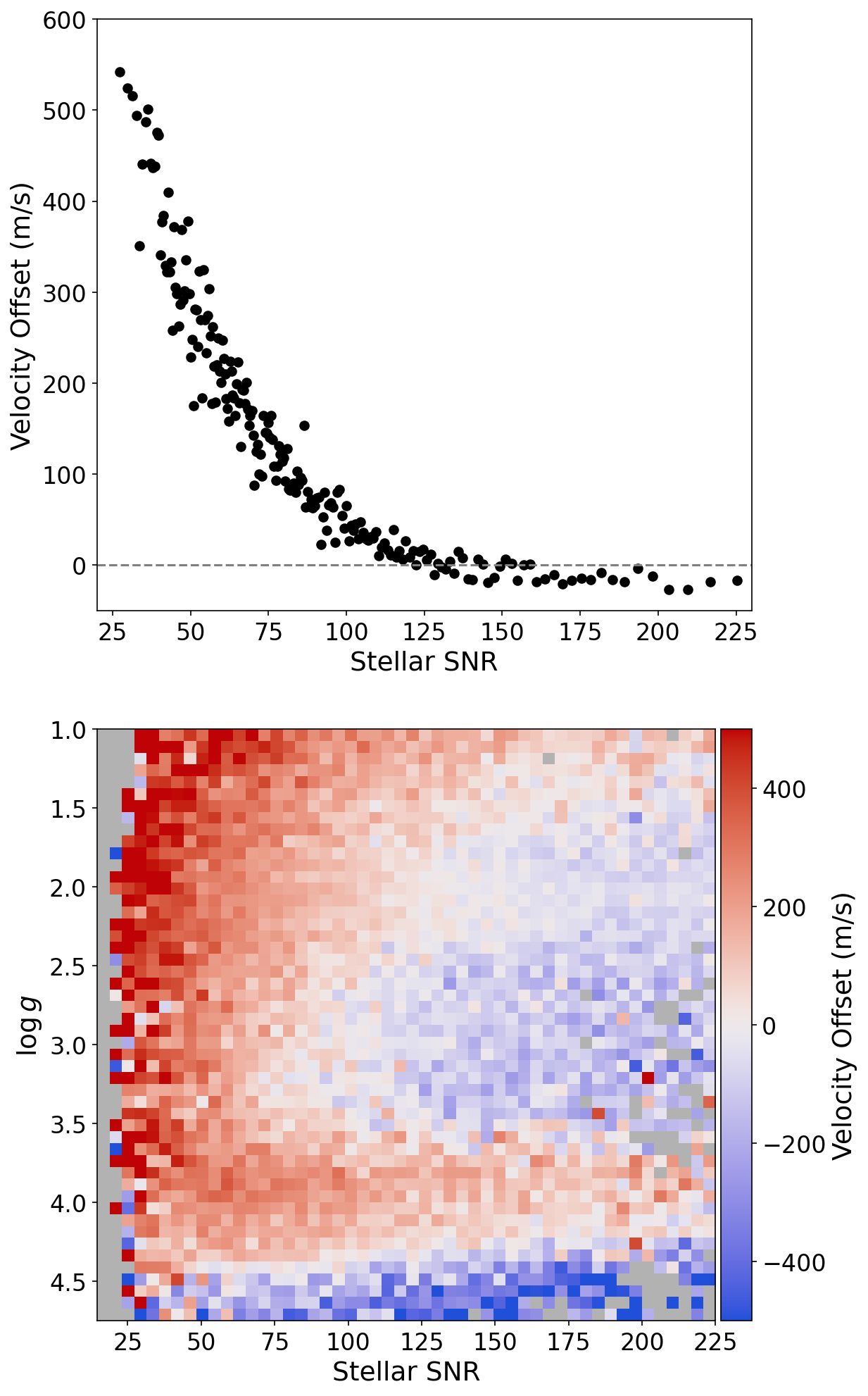}
\caption{\aks{
(Top) Velocity shift in the stellar rest frame as measured from the MADGICS residuals by comparing the magnitude a fit of a Lorentzian derivative to a Lorentzian fit of the stellar component of the spectra. (Bottom) Measurements of the average shift in the core of the Ca triplet lines using Lorentzian fits to Gaia data alone. Zero velocity reference arbitrarily chosen to be the (stellar SNR, $\log g$) $\approx (150, 2)$ bin. Comparison shows MADGICS residuals are sensitive enough to show small velocity shifts below the reported precision in Gaia radial velocities that are present in the data themselves.
}}
\label{fig:velShift}
\end{figure}

\aks{ 

To measure the magnitude of the rest-frame shift indicated by the residuals in Figure 13, we performed a least-squares fit of the residuals with the derivative of a Lorentzian line shape and compared that with the amplitude of a Lorentzian fit to the stellar components for the same stars. Each Ca triplet line was fit independently in a 100 pixel wide window (see Section \ref{sec:dataavil} for more details). The result is plotted as a function of stellar SNR in Figure \ref{fig:velShift} (top) and suggests a velocity shift of $\sim 50$ m/s at SNR $\sim 100$ and $\sim 500$ m/s at SNR $20$. 

In order to confirm this apparent velocity shift is a result of the data, and not a bias introduced by the MADGICS methodology introduced here, we examined the velocity shift more closely in the data from Gaia without using MADGICS at all. We separated the Gaia data (with the same cuts as described in Section \ref{sec:Data}) into 50 bins each in log(g) and SNR. The spectra in each bin were averaged and Lorentzian curves were fit independently to each of the three Ca lines. We then calculated the average velocity shift across all three Ca lines for each bin relative to high SNR giants, (stellar SNR, $\log g$) $\approx (150, 2)$, which is shown in Figure \ref{fig:velShift} (bottom). This analysis shows the same range of velocity offsets (around $\sim 600$ m/s) as fitting the MADGICS residuals, suggesting that this is an effect of the Gaia data propagating through MADGICS.

The asymmetry as a function of SNR could arise from the asymmetric line shape of the Ca triplet, which \cite{AllendePrieto_2013_A_A} showed leads to apparent shifts of a similar order of magnitude in the radial velocity measure as a function of spectral resolution. Given the Gaia RVS resolution and the type-dependent convective shifts $\sim 300$ m/s predicted by \cite{AllendePrieto_2013_A_A}, these residuals are impressive and entirely consistent with the reported precision of Gaia radial velocities. So, the clarity of these small velocity offsets in the MADGICS residuals is a testament to the sensitivity of the method for determining radial velocity measures. In addition, to our knowledge, the strong SNR dependence of the Gaia DR3 RVS spectra Ca triplet alignment at low SNR has not been previously reported.
}

\end{appendices}


\bibliography{GaiaDIB.bib}{}
\bibliographystyle{aasjournal}
\end{document}